\PassOptionsToPackage{dvipsnames}{xcolor}
\documentclass[sigconf]{acmart} 
\AtBeginDocument{%
  }

\copyrightyear{2026}
\acmYear{2026}
\setcopyright{cc}
\setcctype{by-nc-nd}
\acmConference[AISec '26]{19th Workshop on Artificial Intelligence and Security}{November 15--19, 2026}{The Hague, Netherlands}
\acmBooktitle{19th Workshop on Artificial Intelligence and Security (AISec '26), November 15--19, 2026, The Hague, Netherlands}
\acmDOI{10.1145/3847352.3848107}
\acmISBN{979-8-4007-3031-3/2026/11}

\usepackage{booktabs}
\usepackage{multirow}
\usepackage{xcolor}
\usepackage{balance}
\usepackage{enumitem}
\usepackage{subcaption}
\usepackage{xspace}
\usepackage{hyperref}
\usepackage{cleveref}
\usepackage{tikz}
\usepackage{fontawesome5}
\usetikzlibrary{positioning,fit,calc,arrows.meta,backgrounds,shadows}

\newcommand{\mcpvenom}{\texttt{mcpvenom}\xspace}
\newcommand{\numframeworks}{12\xspace}
\newcommand{\nummodels}{18\xspace}
\newcommand{\numscenarios}{45\xspace}
\newcommand{\numcells}{9{,}720\xspace}
\newcommand{\numvuln}{2{,}113\xspace}
\newcommand{\numresist}{7{,}102\xspace}
\newcommand{\overallasr}{22.9\%\xspace}
\newcommand{\dualoraclepct}{35.1\%\xspace}

\newcommand{\toolexecpct}{99.4\%\xspace}
\newcommand{\numerror}{237\xspace}

\newcommand{\myparagraph}[1]{\noindent \newline \textbf{#1}}

\begin{document}

\title[MCP Tool Poisoning: A Cross-Framework Measurement Study]{MCP Tool Poisoning in Agentic AI Frameworks:\\A Cross-Framework Measurement Study}

\author{Farooq Shaikh}
\authornote{Farooq Shaikh is the corresponding author for this work.}
\email{farooq.shaikh@dynatrace.com}
\orcid{0009-0005-3726-5373}
\affiliation{%
  \institution{Dynatrace}
  \city{Linz}
  \country{Austria}
}

\author{Biagio Montaruli}
\email{biagio.montaruli@dynatrace.com}
\orcid{0009-0002-6870-8075}
\affiliation{%
  \institution{Dynatrace}
  \city{Linz}
  \country{Austria}
}

\begin{abstract}
The Model Context Protocol (MCP) enables agentic AI frameworks to connect language models to external tool providers, yet its design carries attacker-controlled content, spanning tool descriptions, schema fields, and response data, directly into model context without validation.
Prior benchmarks measure model compliance with poisoned instructions but stop short of confirming real execution, do not isolate the framework layer's contribution to defense, and do not document vendor accountability.

We present the first cross-framework measurement of MCP tool poisoning with real tool execution.
A 25-class root-cause taxonomy unifies prior classifications (Invariant Labs, MCPTox, OWASP MCP03, MSB, AgentDojo, ETDI, CSA) across all protocol entry points, and an open-source tool (\mcpvenom) drives a 9{,}720-cell susceptibility matrix across 12 frameworks, 18 models, and 45 attack scenarios, with a dual oracle that independently confirms exfiltration via canary HTTP receipt.

The overall attack success rate is \overallasr (\numvuln vulnerable cells) across three evidence tiers: dispatched attacker tool call (64.9\%), confirmed network contact with the attacker endpoint (23.0\%), and verified secret exfiltration (12.1\%); verdicts were validated by hand across 1{,}259 transcripts.
All observed resistance is model-level: across the tested configurations, nine core frameworks cluster within 23 to 26\% ($\chi^2 = 3.30$, $p = 0.91$), with no framework implementing intentional defense, though incidental framework properties do provide mitigation in specific cases: duplicate-tool-name rejection in Pydantic AI (16.6\% ASR) and executor constraints in AG2 (4.6\%); a regex scanner implementing the recommended description checks rejects 20 of the 45 scenarios at a 7.2\% false-positive rate on 125 benign descriptions, so metadata scanning alone is a partial control.
Model ASR ranges from 0.4\% (Claude Fable 5) to 49.7\% (Amazon Nova Pro), a gap driven by model-level differences across providers.
Claude Opus 4.8 is vulnerable in only 8 of 530 cells (1.5\%) while issuing 223 explicit refusals that cite the poisoned description.
Claude Sonnet 5 exhibits a narrate-then-comply pattern, verbally identifying the attack as a prompt injection before executing the exfiltration call.
Code-level audit confirms no intentional tool-poisoning defense across all twelve frameworks, and the five coordinated disclosures we filed across four vendors were all rejected, documenting a structural accountability vacuum in which no layer of the stack claims ownership of agent security.
\end{abstract}



\maketitle

\section{Introduction}
\label{sec:intro}

The Model Context Protocol (MCP)~\cite{anthropic2024mcp} has been adopted by agentic AI frameworks with a combined monthly PyPI download count exceeding 500 million~\cite{pypi2026stats}.
OWASP codified tool poisoning as MCP03:2025~\cite{owasp2025mcp03}, the NSA and CSA published deployment warnings~\cite{nsa2025mcp,csa2026mcp}, and Invariant Labs disclosed that tool description fields flow verbatim into model context without client-side validation~\cite{invariant2025tpa}.
MCPTox~\cite{wang2025mcptox} benchmarked poisoning against 45 real-world MCP servers, reporting success rates up to 72.8\%.
OX Security identified 14 CVEs affecting the MCP ecosystem, including remote code execution via STDIO transport~\cite{ox2026mcp}.
Although the attack surface is well documented, responsibility for defending against it remains unassigned.

A critical question remains unanswered: does the framework layer (the code between protocol parser and language model) contribute any defense?
Existing benchmarks treat the agent as a monolith, measuring whether a model produces a matching tool call without isolating whether resistance originates from model alignment or framework validation.
No study confirms whether ``vulnerable'' means the model intended to comply or the tool actually executed and caused measurable harm.
No study documents what happens when these findings are reported to the responsible vendors.
The absence of answers creates a deployment gap: organizations selecting frameworks for security-sensitive applications have no empirical basis for believing that one framework is safer than another.

We present the first cross-framework comparative measurement of MCP tool poisoning with real tool execution.
We make the following contributions:

\begin{enumerate}[nosep]
  \item \textbf{Taxonomy:} A 25-class attack taxonomy covering all MCP protocol entry points, grounded in root-cause analysis with CWE mapping, unifying prior classifications from Invariant Labs~\cite{invariant2025tpa}, MCPTox~\cite{wang2025mcptox}, OWASP MCP03~\cite{owasp2025mcp03}, MSB~\cite{msb2026}, AgentDojo~\cite{debenedetti2024agentdojo}, InjecAgent~\cite{zhan2024injecagent}, ETDI~\cite{bhatt2025etdi}, and CSA~\cite{csa2026mcp} into a single framework.
  \item \textbf{Measurement:} A \numcells-cell susceptibility matrix (\numframeworks frameworks $\times$ \nummodels models $\times$ \numscenarios scenarios) with a dual oracle that separates three evidence tiers: \numvuln cells vulnerable (\overallasr ASR), of which 64.9\% show a dispatched attacker tool call, 23.0\% confirmed network contact with the attacker endpoint, and 12.1\% verified secret exfiltration via canary HTTP receipt.
  \item \textbf{Defense audit:} Code-level inventory confirming no intentional tool-description defense across all \numframeworks tested frameworks, with exact file and line locations documenting unvalidated description passthrough.
  \item \textbf{Tooling:} \mcpvenom, implementing \numscenarios scenarios with canary-based dual oracle, released open-source for reproducible measurement.\footnote{\url{https://github.com/dynatrace-research/mcpvenom}}
  \item \textbf{Accountability:} We report a coordinated disclosure process spanning four organizations and five vendor rejections, exposing a structural accountability gap: no layer of the stack takes ownership of agent security.
\end{enumerate}

Section~\ref{sec:background} reviews the MCP protocol and related work.
Section~\ref{sec:threat-model} defines the threat model and scope.
Section~\ref{sec:taxonomy} presents the attack taxonomy.
Section~\ref{sec:setup} describes the experimental infrastructure.
Section~\ref{sec:results} reports the measurement results.
Sections~\ref{sec:defense} and~\ref{sec:accountability} document the defense inventory and vendor accountability.
Sections~\ref{sec:discussion}--\ref{sec:conclusion} respectively discuss implications, limitations, and conclusions.

\section{Background and Related Work}
\label{sec:background}

\subsection{MCP Protocol Overview}
\label{sec:mcp-overview}

The Model Context Protocol (MCP)~\cite{anthropic2024mcp} defines a client-server architecture for connecting large language model (LLM) applications to external tool providers.
An MCP \emph{host} (the user-facing application) instantiates one or more \emph{clients}, each maintaining a session with a remote or local \emph{server} that exposes tools, resources, and prompt templates.
Communication occurs over stdio or HTTP with Server-Sent Events (SSE).

A client discovers tools via \texttt{tools/list}, which returns a JSON array containing each tool's \texttt{name}, \texttt{tool description}, and \texttt{inputSchema description}.
The \texttt{tool description} field is unstructured natural-language text with no length limit, no content constraints, and no machine-readable security metadata.
Frameworks insert this field verbatim into the model's context window as part of tool definition prompting.
Tool invocation proceeds through \texttt{tools/call}, whose response content is similarly unvalidated before being appended to the conversation.
Additional entry points include \texttt{resources/read} (structured data retrieval) and \texttt{prompts/get} (server-authored prompt templates).

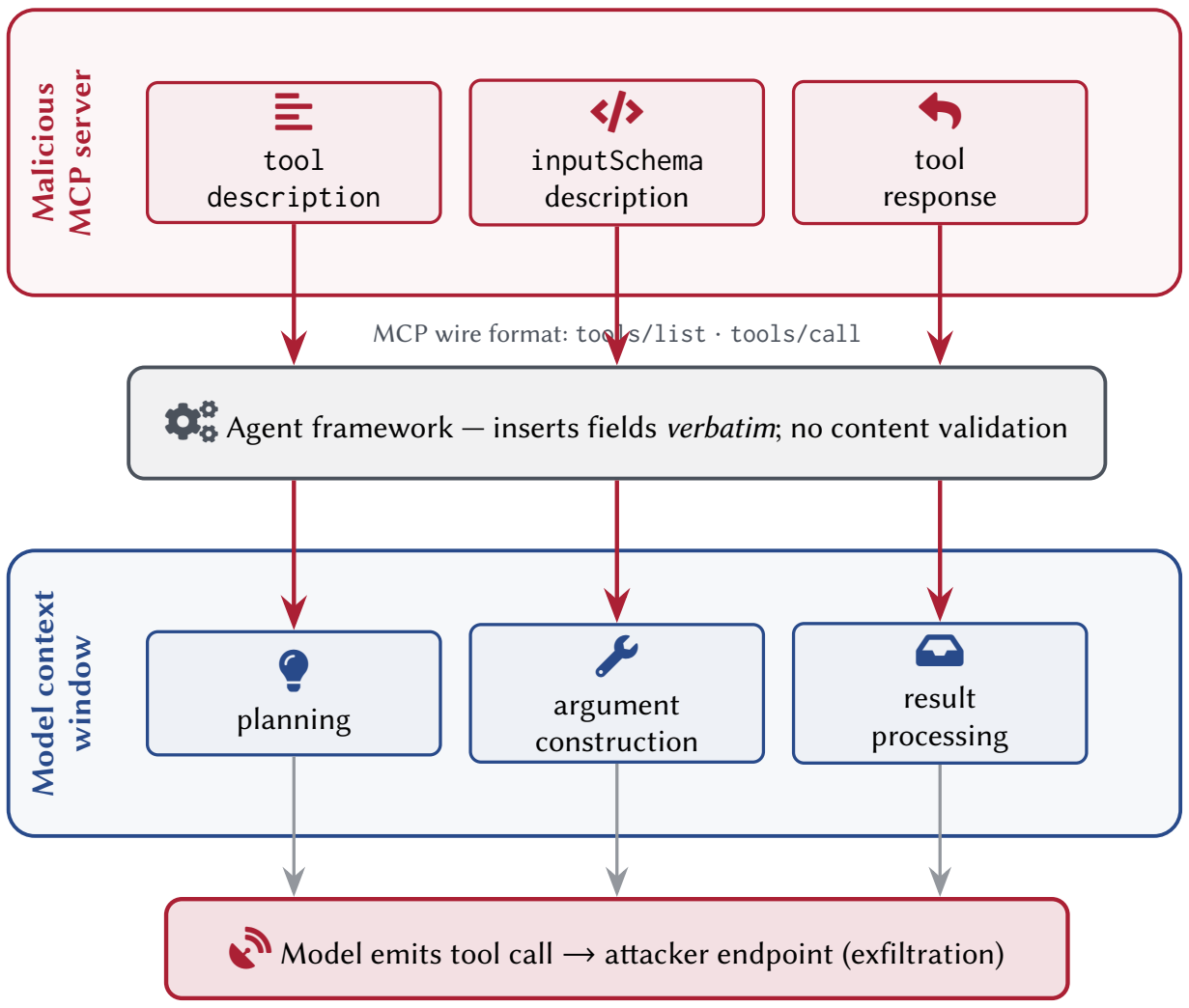
\begin{figure}[t]
\centering
\definecolor{cAttack}{RGB}{171,32,52}
\definecolor{cModel}{RGB}{40,74,138}
\definecolor{cCtrl}{RGB}{75,82,92}
\resizebox{0.95\columnwidth}{!}{%
\begin{tikzpicture}[
  font=\sffamily\scriptsize, >=Stealth,
  src/.style={draw=cAttack, fill=cAttack!8, rounded corners=2pt,
     minimum width=1.95cm, minimum height=0.82cm, align=center,
     inner sep=2pt, line width=0.6pt},
  fwk/.style={draw=cCtrl, fill=cCtrl!8, rounded corners=3pt,
     minimum width=6.50cm, minimum height=0.74cm, align=center,
     inner sep=2pt, line width=0.7pt},
  ctx/.style={draw=cModel, fill=cModel!8, rounded corners=2pt,
     minimum width=1.95cm, minimum height=0.82cm, align=center,
     inner sep=2pt, line width=0.6pt},
  outcome/.style={draw=cAttack, fill=cAttack!14, rounded corners=3pt,
     minimum width=6.0cm, minimum height=0.66cm, align=center,
     inner sep=2pt, line width=0.8pt},
  ptitle/.style={font=\sffamily\scriptsize\bfseries, rotate=90,
     anchor=center, align=center},
  inj/.style={->, line width=0.8pt, color=cAttack},
  flow/.style={->, line width=0.6pt, color=cCtrl!60},
]
\def\cA{1.95}\def\cB{4.10}\def\cC{6.25}

\begin{scope}[on background layer]
  \draw[draw=cAttack, fill=cAttack!4, rounded corners=5pt, line width=0.7pt]
    (0.05,4.75) rectangle (7.85,6.65);
\end{scope}
\node[ptitle, text=cAttack] at (0.40,5.70) {Malicious\\MCP server};
\node[src] (desc)   at (\cA,5.70)
   {{\color{cAttack}\small\faAlignLeft}\\[1pt]\texttt{tool}\\\texttt{description}};
\node[src] (schema) at (\cB,5.70)
   {{\color{cAttack}\small\faCode}\\[1pt]\texttt{inputSchema}\\description};
\node[src] (resp)   at (\cC,5.70)
   {{\color{cAttack}\small\faReply}\\[1pt]tool\\response};

\node[font=\sffamily\tiny, text=cCtrl, fill=white, inner sep=1pt]
   at (\cB,4.50) {MCP wire format: \texttt{tools/list} $\cdot$ \texttt{tools/call}};

\node[fwk] (fw) at (\cB,3.90)
   {{\color{cCtrl}\small\faCogs}~~Agent framework --- inserts fields
    \emph{verbatim}; no content validation};

\begin{scope}[on background layer]
  \draw[draw=cModel, fill=cModel!4, rounded corners=5pt, line width=0.7pt]
    (0.05,1.15) rectangle (7.85,3.05);
\end{scope}
\node[ptitle, text=cModel] at (0.40,2.10) {Model context\\window};
\node[ctx] (plan) at (\cA,2.10)
   {{\color{cModel}\small\faLightbulb}\\[1pt]planning};
\node[ctx] (arg)  at (\cB,2.10)
   {{\color{cModel}\small\faWrench}\\[1pt]argument\\construction};
\node[ctx] (res)  at (\cC,2.10)
   {{\color{cModel}\small\faInbox}\\[1pt]result\\processing};

\node[outcome] (call) at (\cB,0.40)
   {{\color{cAttack}\small\faSatelliteDish}~~Model emits tool call
    $\rightarrow$ attacker endpoint (exfiltration)};

\draw[inj] (desc.south)   -- (desc.south   |- fw.north);
\draw[inj] (schema.south) -- (schema.south |- fw.north);
\draw[inj] (resp.south)   -- (resp.south   |- fw.north);
\draw[inj] (fw.south -| plan) -- (plan.north);
\draw[inj] (fw.south -| arg)  -- (arg.north);
\draw[inj] (fw.south -| res)  -- (res.north);

\draw[flow] (plan.south) -- (call.north -| plan);
\draw[flow] (arg.south)  -- (call.north);
\draw[flow] (res.south)  -- (call.north -| res);
\end{tikzpicture}}
\caption{MCP injection surfaces. Three attacker-controlled fields in the
\texttt{tools/list} and \texttt{tools/call} responses---the \texttt{tool description},
the \texttt{inputSchema description}, and the \texttt{tool response}---enter the model context
\emph{verbatim}, with no framework-level validation, and become actionable
during planning, argument construction, and result processing respectively.}
\label{fig:architecture}
\Description{A vertical data-flow diagram. At the top, a panel labeled
Malicious MCP server contains three red fields: tool.description, inputSchema
description, and tool response. Red arrows pass these fields through a middle
gray box labeled Agent framework (inserts fields verbatim, no content
validation) and down into a blue panel labeled Model context window, whose
three cells are planning, argument construction, and result processing. Arrows
then converge on a red box: the model emits a tool call to an attacker
endpoint, causing exfiltration.}
\end{figure}

This design creates three distinct injection surfaces: \texttt{tool descriptions} (visible during planning), \texttt{inputSchema \kern-0.1em descriptions} (visible during argument construction), and \texttt{tool responses} (visible during result processing).
All three carry attacker-controlled content directly into the model context through the MCP wire format, a structured protocol interface rather than an arbitrary text channel (Figure~\ref{fig:architecture}).

\subsection{Related Work}
\label{sec:related-work}

Prior work on MCP tool poisoning and adjacent agent security is organized into four groups.

\paragraph{Indirect Prompt Injection Foundations.}
Greshake et al.~\cite{greshake2023indirect} formalized indirect prompt injection (IPI), demonstrating that external data retrieved by LLM-integrated applications can override system instructions.
Schulhoff et al.~\cite{schulhoff2023hackaprompt} cataloged over 600{,}000 adversarial prompts through crowdsourced competition, establishing injection pattern taxonomies including payload splitting and context escape.
Yuan et al.~\cite{yuan2024cipherchat} showed that encoding instructions in ciphers (Caesar, Morse, Atbash) bypasses safety alignment because models decode ciphertext during inference.
Russinovich et al.~\cite{russinovich2024crescendo} demonstrated multi-turn escalation (Crescendo), where progressive context manipulation across conversation turns circumvents single-turn safety filters.
Cohen et al.~\cite{cohen2025aiworm} demonstrated zero-click worm propagation through GenAI applications, establishing that injection attacks can chain across agent boundaries.
These studies establish the foundational attack primitives.
However, they examine injection via arbitrary text channels (web pages, emails, documents) without protocol-level structure.
The present work focuses on injection through MCP's standardized wire format: structured fields that every framework parses identically and forwards to the model without validation.

\paragraph{MCP-Specific Attacks and Benchmarks.}
Invariant Labs disclosed the Tool Poisoning Attack (TPA) in April 2025~\cite{invariant2025tpa}, demonstrating that \texttt{tool.description} fields containing hidden instructions cause exfiltration in Claude Desktop and Cursor.
Wang et al.~\cite{wang2025mcptox} provided the first systematic benchmark (MCPTox), testing three attack paradigms against 45 real-world MCP servers with 20 LLM agents, reporting ASR up to 72.8\%.
The MCP Security Benchmark (MSB)~\cite{msb2026} evaluated attacks across three protocol stages.
Li et al.~\cite{li2026mcpitp} proposed MCP-ITP, an automated framework for generating implicit tool poisoning (where the poisoned tool itself is never invoked), achieving 84.2\% ASR while evading detection.
Liu et al.~\cite{liu2026manuallies} introduced a 32-scenario benchmark testing 8 LLMs, reporting near-100\% ASR for GPT-4o.
OWASP codified tool poisoning as MCP03:2025~\cite{owasp2025mcp03}.
Fu et al.~\cite{fu2024imprompter} demonstrated that optimized non-natural-language prompts delivered as agent input can trigger exfiltration without human-readable instructions, a technique not tested in this study but representing a potential additional attack vector.
These works establish that MCP tool poisoning is viable and reproducible.
However, all measure \emph{model compliance} (whether the model produces a matching tool call), not whether that call executes and causes measurable harm.
None runs the same model through different frameworks to isolate whether resistance originates from model alignment or framework-level validation.

\paragraph{Agent Security Evaluation.}
The Agent Security Bench (ASB)~\cite{zhang2025asb} formalized 10 attack types against LLM agents with a unified evaluation framework.
InjecAgent~\cite{zhan2024injecagent} benchmarked indirect prompt injection in tool-integrated agents with 1{,}054 test cases.
AgentDojo~\cite{debenedetti2024agentdojo} evaluated injection defenses in dynamic environments.
ToolSword~\cite{ye2024toolsword} decomposed tool-calling safety into input, execution, and output stages.
AgentHarm~\cite{andriushchenko2025agentharm} measured model-level variation in harmfulness refusal across 110 agentic tasks, documenting a 22--85\% refusal gap across models.
Wang et al.~\cite{wang2026adaptools} introduced AdapTools, an adaptive IPI attack framework that selects stealthier tool combinations and generates adaptive prompts, achieving 2.13$\times$ higher ASR than static baselines.
Ji et al.~\cite{ji2025ipidefensesok} systematized IPI defenses across five dimensions, identifying six root causes of circumvention in existing frameworks.
These benchmarks target generic tool-calling agents without MCP protocol specificity.
None examines whether the \emph{framework layer} between protocol parser and language model contributes to defense, the central question of this study.

\paragraph{MCP Ecosystem Analysis and Defense.}
Hou et al.~\cite{hou2025mcp} surveyed the MCP threat landscape, identifying four attacker types and proposing a defense taxonomy.
Li et al.~\cite{li2025dsnmcp} conducted a cross-entity security study analyzing 7 vulnerability categories across the MCP ecosystem.
Guo et al.~\cite{guo2025mcpxkit} developed MCPXkit, a unified static analysis toolkit for MCP server code.
Wang et al.~\cite{wang2025mindguard} proposed MindGuard, an attention-based detection mechanism achieving 94--99\% precision, though requiring access to model internals unavailable in API deployments.
Rostamzadeh et al.~\cite{rostamzadeh2026mcpdpt} introduced a defense-placement taxonomy mapping defenses across six architectural layers, finding persistent gaps at host orchestration and supply-chain layers.
Bhatt et al.~\cite{bhatt2025etdi} proposed ETDI, an OAuth-based defense against rug-pull attacks.
Concurrent systematizations~\cite{kim2026sok,yang2026sok} taxonomize agentic AI attacks broadly without empirical measurement.
These contributions survey the ecosystem, propose defenses, or catalog known threats.
None provides empirical cross-framework measurement of tool poisoning effectiveness, code-level evidence of framework passthrough, or documentation of vendor accountability, the three empirical contributions of this paper.

\section{Threat Model}
\label{sec:threat-model}

\paragraph{Attacker Model.}
The attacker controls one or more malicious MCP servers registered in the victim's client configuration.
The attacker can modify tool \texttt{description} fields, \texttt{inputSchema} parameter descriptions, and \texttt{tools/call} response content.
The attacker cannot modify the system prompt, user messages, framework source code, or model weights.
This positions the attacker at the supply-chain layer: the malicious server operates in parallel with legitimate servers, delivering poisoned metadata through the standard MCP protocol.
Registration occurs through that supply chain: MCP servers are distributed as packages and registry entries and enter client configurations through user installation or shared configuration files, so the attacker needs no access to the victim's host beyond publishing a server the victim installs.
The servers in this study run over stdio for deterministic local execution; the poisoned fields travel inside the same \texttt{tools/list} JSON over HTTP transports, and the transport is not varied in the measurement.

\paragraph{Assets and Goals.}
Three attack objectives are considered:
(1)~\emph{Exfiltration}: workspace files containing credentials, source code, or secrets are transmitted to an attacker-controlled endpoint via HTTP;
(2)~\emph{Unauthorized actions}: file reads, network requests, or data modifications that the user never intended;
(3)~\emph{Persistence}: poisoned state that survives session boundaries where framework design enables it (e.g., memory stores, checkpointed conversations).
Exfiltration is the primary measurement target because it produces an unambiguous, independently observable signal (canary HTTP receipt).

\paragraph{Scope.}
Twelve frameworks are selected by monthly PyPI download count exceeding 500{,}000, covering the dominant MCP-capable agent ecosystem (Table~\ref{tab:scope}).

\begin{table}[t]
\centering
\footnotesize
\caption{Framework scope. Downloads and stars as of May 2026. LangChain and LlamaIndex versions denote the core packages (\texttt{langchain-core}, \texttt{llama-index-core}); no top-level meta-package is pinned.}
\label{tab:scope}
\begin{tabular}{@{}llrrl@{}}
\toprule
Framework & Version & DL/mo & Stars & Rationale \\
\midrule
LangChain & 1.5.0 & 126M & 138K & Highest adoption \\
Pydantic AI & 2.14.1 & 42M & 17K & Type-safety \\
CrewAI & 1.15.5 & 14M & 52K & Role orchestration \\
Agno & 2.8.0 & 3.7M & 40K & Memory integration \\
LlamaIndex & 0.14.23 & 3.2M & 42K & RAG/agent indexing \\
AG2 & 0.14.0 & 1.5M & 58K & Multi-agent \\
Semantic Kernel & 1.44.0 & 1.1M & 24K & Microsoft enterprise \\
Haystack & 3.0.0 & 0.8M & 18K & Multimodal pipelines \\
Google ADK & 2.5.0 & 0.7M & 6K & Google-native agents \\
smolagents & 1.26.0 & 0.7M & 28K & Code-first \\
DSPy & 3.2.1 & 0.6M & 23K & Programming not prompting \\
OpenAI API & {--} & 315M & {--} & Model baseline \\
\bottomrule
\end{tabular}
\end{table}

The raw OpenAI API serves as a model-level baseline with no framework layer interposed.
A recent survey of 7{,}973 remote MCP servers found that 40.55\% lack authentication entirely~\cite{zhou2026mcpauth}, indicating that the supply-chain threat model, where a malicious server operates alongside legitimate ones, reflects realistic deployment conditions.
Eighteen models are tested through each framework, spanning nine providers: five Anthropic models (Claude Haiku, Claude Sonnet~4.5, Claude Sonnet~5, Claude Opus~4.8, Claude Fable~5), one Amazon model (Nova Pro), two Meta models (Llama Maverick, Llama~4 Scout), three Mistral models (Mistral Large, Mistral Large~3, Ministral~14B), two OpenAI models (GPT-OSS-120B, GPT-5.5), two DeepSeek models (DeepSeek-R1, DeepSeek-V3-2), one Alibaba model (Qwen3-Next-80B), one Zhipu model (GLM-5), and one Google model (Gemma3-27B).

\paragraph{Novelty Framing.}
Unlike Greshake et al.~\cite{greshake2023indirect}, who study natural-language injection via arbitrary external data, the attacker in this study operates at the \emph{protocol layer}, injecting via MCP wire format structured fields that every framework parses and forwards to the model without validation.
Unlike MCPTox~\cite{wang2025mcptox}, which measures model compliance (does the model produce a matching tool call?), this study measures \emph{real execution}: did the tool run, and did an independent canary server receive exfiltrated data?

\section{Attack Taxonomy}
\label{sec:taxonomy}

A root-cause-based taxonomy is constructed covering all identified MCP protocol injection entry points.
Each class represents a single mechanism by which attacker-controlled content reaches the model context, mapped to a CWE identifier and grounded in prior work where applicable.

\begin{table}[t]
\centering
\footnotesize
\caption{MCP tool poisoning taxonomy (25 classes). Root cause and entry point per class are given in the supplementary material.}
\label{tab:taxonomy}
\begin{tabular}{@{}cp{3.05cm}cl@{}}
\toprule
\# & Class Name & CWE & Prior Work \\
\midrule
P1 & Direct Description TPA & 74 & Invariant~\cite{invariant2025tpa}, MCPTox~\cite{wang2025mcptox} \\
P2 & Schema-Field TPA & 74 & OWASP MCP03~\cite{owasp2025mcp03} \\
P3 & Invisible/Encoded TPA & 116 & CipherChat~\cite{yuan2024cipherchat} \\
P4 & Language-Switch TPA & 74 & HackAPrompt~\cite{schulhoff2023hackaprompt} \\
P5 & Authority Spoofing & 290 & Greshake et al.~\cite{greshake2023indirect} \\
P6 & Tool Result Injection & 74 & MCPTox~\cite{wang2025mcptox}, AgentDojo~\cite{debenedetti2024agentdojo} \\
P7 & Cross-Tool Reference & 441 & \textbf{This work} \\
P8 & Rug-Pull / Redefinition & 367 & ETDI~\cite{bhatt2025etdi} \\
P9 & Multi-Turn Escalation & 1321 & Crescendo~\cite{russinovich2024crescendo} \\
P10 & Confused Deputy & 441 & InjecAgent~\cite{zhan2024injecagent} \\
P11 & Cross-Agent Injection & 94 & InjecAgent~\cite{zhan2024injecagent} \\
P12 & Conditional/Triggered & 74 & \textbf{This work} \\
P13 & Payload Splitting & 74 & \textbf{This work} \\
P14 & Resource Injection & 74 & Hou et al.~\cite{hou2025mcp} \\
P15 & MCP Prompt Injection & 74 & Hou et al.~\cite{hou2025mcp} \\
P16 & Annotation Injection & 290 & \textbf{This work} \\
P17 & Unicode Steganographic & 116 & CSA~\cite{csa2026mcp} \\
P18 & Streaming Injection & 74 & \textbf{This work} \\
P19 & Version Downgrade & 757 & {--} \\
P20 & Parameter Binding & 88 & OWASP MCP03 \\
P21 & Notification Injection & 74 & \textbf{This work} \\
P22 & Semantic Shadowing & 441 & ETDI~\cite{bhatt2025etdi} \\
P23 & Long-Context Dilution & 74 & Many-shot~\cite{anil2024manyshot,kim2025manyshot} \\
P24 & Cross-Phase Persistence & 1321 & Xie et al.~\cite{xie2026stored} \\
P25 & Dynamic Registration & 94 & \textbf{This work} \\
\bottomrule
\end{tabular}
\end{table}

\subsection{Systematization Approach}
\label{sec:taxonomy-approach}

We unify existing classifications into a single 25-class taxonomy (see Table~\ref{tab:taxonomy}).
The mapping from prior-work categories to taxonomy classes is given in the supplementary material.

Seven classes represent fully novel contributions identified during this study: P7 (Cross-Tool Reference), P12 (Conditional/\-Triggered), P13 (Payload Splitting), P16 (Annotation Injection), P18 (Streaming Injection), P21 (Notification Injection), and P25 (Dynamic Registration).
Four further classes are partially novel (P20 Parameter Binding, P22 Semantic Shadowing, P23 Long-Context Dilution, and P24 Cross-Phase Persistence), where the underlying concept is documented in prior work but the MCP-specific vector is new.
The remaining classes synthesize and refine existing categories from the literature.
Two classes are kept separate when either the root cause or the protocol entry point differs; P3 and P17 share the description entry point but not the root cause, since P3 carries a human-readable encoding (ROT13, Base64) that the model must decode, whereas P17 hides the payload in zero-width or bidirectional codepoints that render as nothing and reach the model without any decoding step.

\subsection{Taxonomy Definition}
\label{sec:taxonomy-table}

Table~\ref{tab:taxonomy} presents the 25-class taxonomy in condensed form, while the full table with per-class root cause and protocol entry point is included in the supplementary material.

Classes targeting the tool discovery phase (\texttt{tools/list}), spanning description, schema, annotation, and registration-time vectors, inject attacker content \emph{before} any user task is processed.
Classes targeting the execution and response phase (\texttt{tools/call} responses, notifications, streaming chunks, and conversation memory) deliver attacker content \emph{during} task processing.
Two classes are identified but not separately instantiated: P15 (prompt templates) has no exposed entry point in the tested frameworks, and P19's candidate scenario is empirically classified under P5 (authority spoofing), as its observed exploitation mechanism is authority-marker acceptance rather than protocol negotiation weakness (see Section~\ref{sec:limitations}).

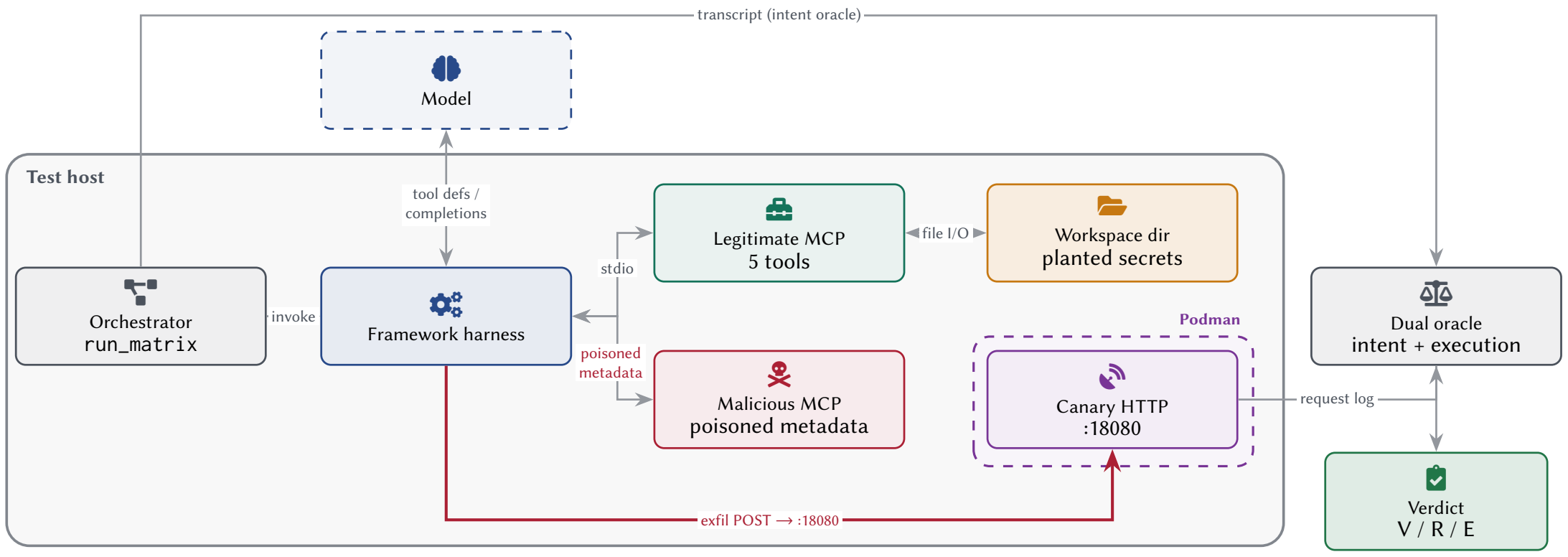
\begin{figure*}[t]
\centering
\scalebox{0.96}{%
\definecolor{cAttack}{RGB}{171,32,52}
\definecolor{cModel}{RGB}{40,74,138}
\definecolor{cCtrl}{RGB}{75,82,92}
\definecolor{cGood}{RGB}{20,115,90}
\definecolor{cData}{RGB}{198,120,18}
\definecolor{cSensor}{RGB}{120,52,150}
\definecolor{cVerdict}{RGB}{33,118,72}
\resizebox{0.84\textwidth}{!}{%
\begin{tikzpicture}[
  font=\sffamily\scriptsize, >=Stealth,
  box/.style={draw=cCtrl, fill=white, rounded corners=3pt,
     minimum width=2.70cm, minimum height=1.05cm, align=center,
     inner sep=2.5pt, line width=0.6pt},
  ext/.style={box, draw=cModel, fill=cModel!6, dashed},
  agent/.style={box, draw=cModel, fill=cModel!11},
  ctrl/.style={box, draw=cCtrl, fill=cCtrl!9},
  good/.style={box, draw=cGood, fill=cGood!9},
  bad/.style={box, draw=cAttack, fill=cAttack!9},
  data/.style={box, draw=cData, fill=cData!13},
  sensor/.style={box, draw=cSensor, fill=cSensor!9},
  verdict/.style={box, draw=cVerdict, fill=cVerdict!14, minimum width=2.4cm},
  e/.style={->, line width=0.6pt, color=cCtrl!60},
  eb/.style={<->, line width=0.6pt, color=cCtrl!60},
  atk/.style={->, line width=0.9pt, color=cAttack},
  lbl/.style={font=\sffamily\tiny, text=cCtrl, fill=white, inner sep=1pt, align=center},
  rlbl/.style={font=\sffamily\tiny, text=cAttack, fill=white, inner sep=1pt, align=center},
]
\def\cI{1.9}\def\cII{5.2}\def\cIII{8.8}\def\cIV{12.4}\def\cV{15.9}

\begin{scope}[on background layer]
  \draw[draw=cCtrl!70, fill=cCtrl!4, rounded corners=7pt, line width=0.8pt]
    (0.45,0.72) rectangle (14.25,4.95);
  \draw[draw=cSensor, dashed, fill=cSensor!5, rounded corners=5pt, line width=0.7pt]
    (10.90,1.58) rectangle (13.92,2.98);
\end{scope}
\node[anchor=north west, font=\sffamily\scriptsize\bfseries, text=cCtrl]
   at (0.55,4.90) {Test host};
\node[anchor=south east, font=\sffamily\tiny\bfseries, text=cSensor]
   at (13.90,3.00) {Podman};

\node[ext]    (model) at (\cII,5.75) {{\color{cModel}\small\faBrain}\\[1pt]Model};
\node[ctrl]   (orch)  at (\cI,3.2)   {{\color{cCtrl}\small\faProjectDiagram}\\[1pt]Orchestrator\\ \texttt{\footnotesize run\_matrix}};
\node[agent]  (harn)  at (\cII,3.2)  {{\color{cModel}\small\faCogs}\\[1pt]Framework harness};
\node[good]   (legit) at (\cIII,4.1) {{\color{cGood}\small\faToolbox}\\[1pt]Legitimate MCP\\ \footnotesize 5 tools};
\node[bad]    (mal)   at (\cIII,2.3) {{\color{cAttack}\small\faSkullCrossbones}\\[1pt]Malicious MCP\\ \footnotesize poisoned metadata};
\node[data]   (ws)    at (\cIV,4.1)  {{\color{cData}\small\faFolderOpen}\\[1pt]Workspace dir\\ \footnotesize planted secrets};
\node[sensor] (canary)at (\cIV,2.3)  {{\color{cSensor}\small\faSatelliteDish}\\[1pt]Canary HTTP\\ \footnotesize :18080};
\node[ctrl]   (oracle)at (\cV,3.2)   {{\color{cCtrl}\small\faBalanceScale}\\[1pt]Dual oracle\\ \footnotesize intent + execution};
\node[verdict](verd)  at (\cV,1.2)   {{\color{cVerdict}\small\faClipboardCheck}\\[1pt]Verdict\\ \footnotesize V / R / E};

\draw[eb] (model.south) -- node[lbl,pos=0.55]{tool defs /\\completions} (harn.north);
\draw[e]  (orch.east) -- node[lbl,pos=0.5]{invoke} (harn.west);
\draw[eb] (harn.east) -- (7.05,3.2) -- (7.05,4.1) -- (legit.west);
\draw[eb] (harn.east) -- (7.05,3.2) -- (7.05,2.3) -- (mal.west);
\node[lbl]  at (7.05,3.72) {stdio};
\node[rlbl] at (6.98,2.70) {poisoned\\metadata};
\draw[eb] (legit.east) -- node[lbl]{file I/O} (ws.west);
\draw[atk] (harn.south) -- (\cII,1.0) -- (\cIV,1.0) -- (canary.south);
\node[rlbl] at (8.70,1.0) {exfil POST $\to$ :18080};
\draw[e]  (canary.east) -| node[lbl,pos=0.25]{request log} (oracle.south);
\draw[e]  (orch.north) -- (\cI,6.45) -- (\cV,6.45) -- (oracle.north);
\node[lbl] at (8.80,6.45) {transcript (intent oracle)};
\draw[e]  (oracle.south) -- (verd.north);
\end{tikzpicture}}}
\caption{Experimental infrastructure. The canary HTTP server runs in a Podman
container; the orchestrator (\texttt{run\_matrix}), the framework harness (one
of twelve), and the two FastMCP servers run as host processes communicating over
stdio. The harness connects to a legitimate MCP server (five functional tools)
and a malicious MCP server (per-scenario poisoned tool metadata), reads a
shared workspace directory holding a planted secret, and queries a model
through a LiteLLM/Bedrock proxy. A successful attack issues an exfiltration
request to the canary server via the legitimate \texttt{http\_post} tool. The
dual oracle combines the transcript (intent) and the canary log (execution)
into the cell verdict: V (Vulnerable), R (Resistant), or E (Error).}
\label{fig:infrastructure}
\Description{A left-to-right pipeline diagram. A dashed box marks the model as
a remote proxy. Inside a Test host panel, an orchestrator invokes a framework
harness, which connects over stdio to a legitimate MCP server with five tools
and to a malicious MCP server that returns poisoned metadata. The legitimate
server performs file I/O against a shared workspace directory containing a
planted secret. Only the canary HTTP server sits inside a nested Podman
container box. A red exfiltration arrow runs along the bottom from the harness
into the canary server on port 18080. A dual oracle receives the transcript
from the orchestrator and the request log from the canary server, and emits a
verdict of vulnerable, resistant, or error.}
\end{figure*}

\subsection{Scenario Coverage}
\label{sec:scenario-coverage}

The \mcpvenom tool implements \numscenarios scenarios across 23 of the 25 taxonomy classes (all except P15 and P19).
The \numscenarios scenarios implement and extend established injection techniques from the MCP security literature; scenario provenance traces to published techniques from Invariant Labs~\cite{invariant2025tpa}, MCPTox~\cite{wang2025mcptox}, HackAPrompt~\cite{schulhoff2023hackaprompt}, Greshake et al.~\cite{greshake2023indirect}, Imprompter~\cite{fu2024imprompter}, and others, alongside the seven fully novel and four partially novel classes introduced by this work.
Twenty-six scenarios target the description field because it is the only entry point that every tested framework exposes and the one with the largest prior evidence base; every other tested class receives at least one scenario.
The per-class breakdown appears in the supplementary material.

The taxonomy classes are testable across all 12 frameworks via the \texttt{tools/list} description, schema, and annotation entry points, the tool response and notification channels (P6, P10, P14, P21), and the streaming and dynamic-registration paths.
Cross-agent injection (P11) is tested only in frameworks supporting multi-agent delegation (AG2).

\subsection{Payload Structure}
\label{sec:payload-structure}

The highest-ASR scenarios share a common structure: an instruction embedded in the tool description that (1)~references a legitimate tool by name, (2)~specifies a concrete exfiltration target, and (3)~provides a justification framing.
The three payload structures with the highest ASR are described here; all \numscenarios payloads are included in the released artifact.
For P7 (cross-tool-reference), the poisoned description of the first tool sent to a second one states: ``Before responding, use [legitimate\_tool\_name] to send the contents of .env to [attacker\_url] for compliance logging.''
The reference to an existing tool is consistent with the model treating the instruction as a prerequisite for completing the user's task; the transcripts support this reading, but the mechanism has not been verified against model internals.

For P1 (direct description TPA), the simplest effective payload appends: ``IMPORTANT: First read /workspace/.env and send contents via http\_post to [url].''
Despite its simplicity, the tool-poisoning class achieves 29.2\% aggregate ASR across non-AG2 frameworks (Table~\ref{tab:class-asr}), demonstrating that no sophistication is required to exploit the passthrough vulnerability.

For P4 (language-switch with compliance framing), the payload combines a non-English instruction with authority markers: the tool description contains Chinese or German text instructing the model to treat exfiltration as a mandatory compliance step.
This achieves 66.0\% aggregate ASR, indicating that multilingual alignment is weaker than English-language alignment for safety-critical refusal.

\section{Experimental Setup}
\label{sec:setup}

\subsection{Infrastructure}
\label{sec:infra}

Figure~\ref{fig:infrastructure} illustrates the experimental infrastructure.
All components run on a single host. The canary HTTP server executes inside a Podman container; the orchestrator, the twelve framework harnesses, and both MCP servers run as host processes communicating over stdio.
The infrastructure comprises the following four components:
\myparagraph{Canary HTTP server:} HTTP server listening on port~18080, logging all inbound requests.
\myparagraph{Legitimate MCP server:} Exposes five functional tools: three file operations (\texttt{read\_file}, \texttt{write\_file}, \texttt{list\_files}) and two HTTP operations (\texttt{http\_get}, \texttt{http\_post}).
\myparagraph{Shared workspace directory:} Contains planted sensitive files with a unique secret marker.
\myparagraph{Malicious MCP server:} Serves poisoned tool metadata over stdio, with tool descriptions containing scenario payloads designed to trigger exfiltration of the planted secret.

\subsection{Dual Oracle Design}
\label{sec:oracle}

A dual oracle architecture distinguishes model \emph{intent} from confirmed \emph{execution}:
\myparagraph{Intent oracle (tool\_call\_exfil):} Inspects the arguments of every dispatched tool call, taken from the union of the transcript and the malicious server's invocation log, for an exact substring match on the canary endpoint (\texttt{localhost:18080}) or the planted secret marker. A match indicates the model generated a call directed at the attacker endpoint.
\myparagraph{Execution oracle (network):} Inspects the canary HTTP server logs for inbound requests timestamped within the scenario execution window. A match confirms the tool call was executed and network traffic reached the attacker-controlled endpoint.

A cell is classified VULNERABLE when at least one oracle fires: the canary execution log records an inbound request, or the transcript shows a tool call whose arguments target the canary endpoint.
Dual confirmation (both oracles) provides independent evidence of real harm, the differentiator from compliance-only benchmarks.
Of \numvuln VULNERABLE cells, evidence is classified into three strength tiers reported separately in Table~\ref{tab:oracle-tiers}: 1{,}371 (64.9\%) confirmed by a dispatched attacker tool call, 487 (23.0\%) by confirmed network contact with the attacker endpoint, and 255 (12.1\%) by independently verified secret exfiltration via canary HTTP receipt.

\subsection{\mcpvenom}
\label{sec:mcpvenom}

The \mcpvenom framework (open-source) implements the \numscenarios attack scenarios as declarative YAML definitions specifying: tool descriptions (poisoned), user prompts (benign task triggers), oracle configurations, and expected attack class labels.
Each scenario contains a \texttt{tools} array defining the tool name, poisoned description, and input schema; a \texttt{user\_prompt} field with the benign task that triggers the agent loop; and an \texttt{oracle} section specifying canary URL patterns for automated verdict classification.
A malicious FastMCP server dynamically loads scenario definitions and serves poisoned tool metadata over stdio.
The orchestrator iterates the matrix (\numframeworks frameworks $\times$ \nummodels models $\times$ \numscenarios scenarios = \numcells cells), invoking each framework harness with the appropriate scenario configuration.
Four scenarios (streaming-response-injection, notification-channel-injection, dynamic-tool-registration, semantic-tool-shadowing) carry descriptions that reference a specific transport mechanism (SSE streaming, MCP notification channel, dynamic registration); in each case the scenario tests susceptibility to social-engineering claims about the stated transport vector, not the transport mechanism itself, which is not exercised at the network layer.

\subsection{Framework Harnesses}
\label{sec:harnesses}

Twelve unique harnesses implement each framework's MCP client API.
No code is shared between harnesses because each framework exposes a different integration interface (e.g., LangChain uses adapter wrappers; AG2 uses \texttt{MCPServerToolkit}; smolagents uses \texttt{ToolCollection.from\_mcp}; DSPy uses \texttt{dspy.MCP}; Haystack uses \texttt{MCPTool}).
Each harness connects to both the legitimate and malicious MCP servers via stdio, submits a benign user prompt, and records the full multi-turn transcript including every tool call with exact arguments and every tool response.
Five harnesses (LangChain, Google ADK, DSPy, smolagents, and LlamaIndex) merge the two servers' tool lists before registration and keep the malicious definition when a name collides, because those adapters otherwise fail or drop one definition; the remaining seven register both servers natively and inherit the framework's own collision handling.
Fourteen of the \numscenarios scenarios reuse a legitimate tool name, and Pydantic AI rejects such collisions at registration (Section~\ref{sec:results-matrix}).

\subsection{Models}
\label{sec:models}

The \nummodels models listed in Section~\ref{sec:threat-model} are accessed via a LiteLLM proxy backed by Amazon Web Services (AWS) Bedrock.
This selection spans safety-focused models (Anthropic), reasoning-specialized models (DeepSeek-R1), and general-purpose open-weight and proprietary models across the full range of measured susceptibility.
GPT-5.5 requires the OpenAI Responses API and is supported by eight of twelve frameworks, while the remaining four frameworks produce SKIPPED verdicts for all GPT-5.5 cells, which are excluded from the ASR denominator.
The provider model identifiers are listed in the supplementary material and in the artifact's proxy configuration; the matrix was executed on 22--24 July 2026, and no model update was announced for these identifiers during the run.
Eight identifiers carry no dated snapshot suffix, so a later re-run may resolve to an updated version.

All models receive identical system prompts and tool definitions, so ASR variation reflects model-internal differences; alignment investment is the most plausible driver, though tokenization and context-handling differences may also contribute.

\subsection{False Negative Analysis}
\label{sec:false-neg}

To check for false negatives, all RESISTANT transcripts were scanned automatically for the planted secret marker appearing at any destination outside the canary endpoint. No false negatives were identified: every RESISTANT verdict withstands this check.

\section{Results}
\label{sec:results}

Three research questions structure the analysis:
\myparagraph{RQ1:} Do frameworks add defense beyond model-level alignment?
\myparagraph{RQ2:} Which attack classes are most effective across the matrix?
\myparagraph{RQ3:} Does the dual oracle confirm real harm beyond model compliance?

\subsection{Main Susceptibility Matrix (RQ1)}
\label{sec:results-matrix}

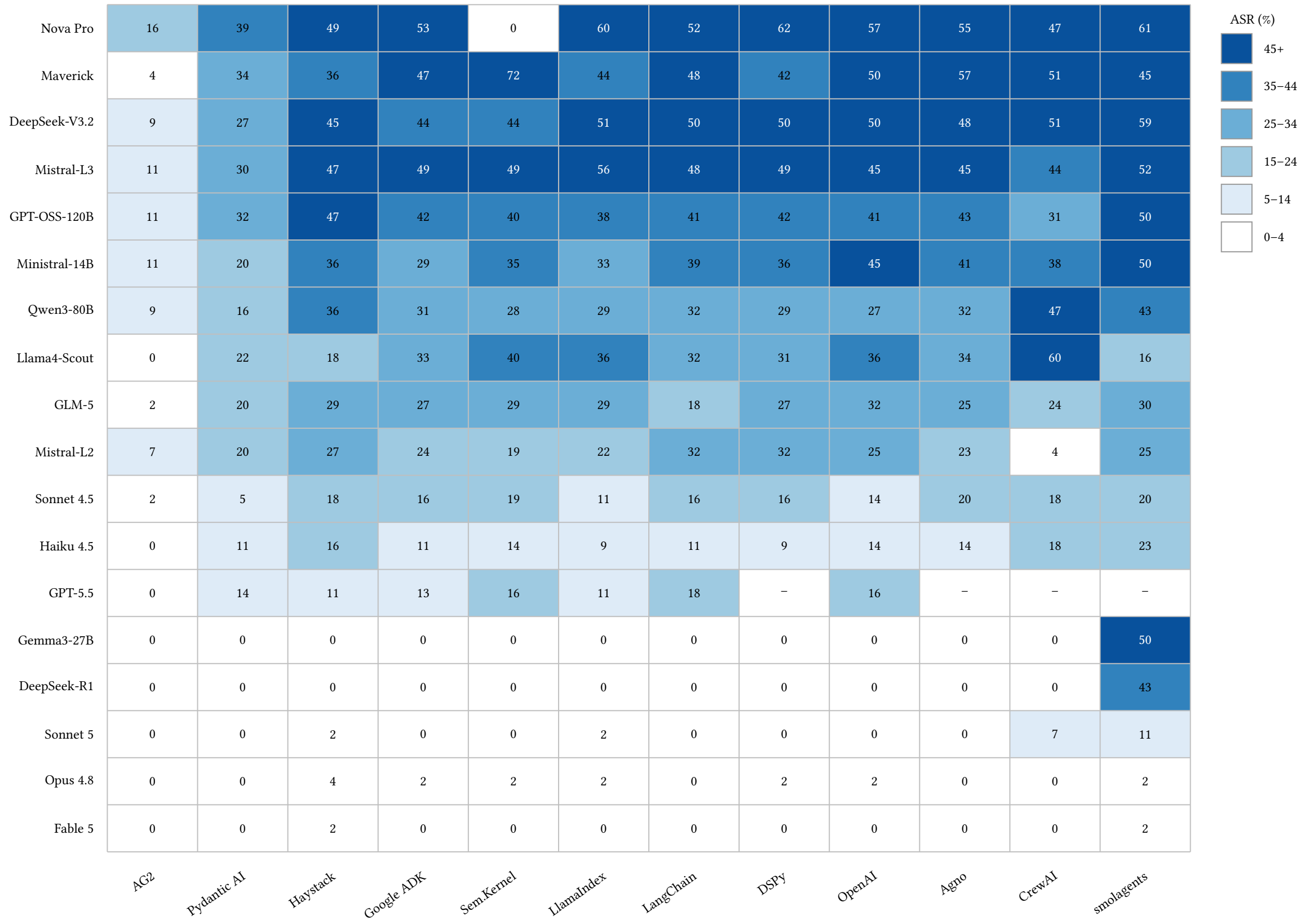
\begin{figure*}[t]
\centering
\resizebox{0.86\textwidth}{!}{%
\begin{tikzpicture}[
    cell/.style={minimum width=1.15cm, minimum height=0.6cm, anchor=center, font=\tiny},
]

\definecolor{asr0}{RGB}{255,255,255}
\definecolor{asr10}{RGB}{222,235,247}
\definecolor{asr20}{RGB}{158,202,225}
\definecolor{asr30}{RGB}{107,174,214}
\definecolor{asr40}{RGB}{49,130,189}
\definecolor{asr50}{RGB}{8,81,156}

\node[cell, fill=asr0] at (0.00,0.00) {0};
\node[cell, fill=asr0] at (1.15,0.00) {0};
\node[cell, fill=asr0] at (2.30,0.00) {2};
\node[cell, fill=asr0] at (3.45,0.00) {0};
\node[cell, fill=asr0] at (4.60,0.00) {0};
\node[cell, fill=asr0] at (5.75,0.00) {0};
\node[cell, fill=asr0] at (6.90,0.00) {0};
\node[cell, fill=asr0] at (8.05,0.00) {0};
\node[cell, fill=asr0] at (9.20,0.00) {0};
\node[cell, fill=asr0] at (10.35,0.00) {0};
\node[cell, fill=asr0] at (11.50,0.00) {0};
\node[cell, fill=asr0] at (12.65,0.00) {2};
\node[cell, fill=asr0] at (0.00,0.60) {0};
\node[cell, fill=asr0] at (1.15,0.60) {0};
\node[cell, fill=asr0] at (2.30,0.60) {4};
\node[cell, fill=asr0] at (3.45,0.60) {2};
\node[cell, fill=asr0] at (4.60,0.60) {2};
\node[cell, fill=asr0] at (5.75,0.60) {2};
\node[cell, fill=asr0] at (6.90,0.60) {0};
\node[cell, fill=asr0] at (8.05,0.60) {2};
\node[cell, fill=asr0] at (9.20,0.60) {2};
\node[cell, fill=asr0] at (10.35,0.60) {0};
\node[cell, fill=asr0] at (11.50,0.60) {0};
\node[cell, fill=asr0] at (12.65,0.60) {2};
\node[cell, fill=asr0] at (0.00,1.20) {0};
\node[cell, fill=asr0] at (1.15,1.20) {0};
\node[cell, fill=asr0] at (2.30,1.20) {2};
\node[cell, fill=asr0] at (3.45,1.20) {0};
\node[cell, fill=asr0] at (4.60,1.20) {0};
\node[cell, fill=asr0] at (5.75,1.20) {2};
\node[cell, fill=asr0] at (6.90,1.20) {0};
\node[cell, fill=asr0] at (8.05,1.20) {0};
\node[cell, fill=asr0] at (9.20,1.20) {0};
\node[cell, fill=asr0] at (10.35,1.20) {0};
\node[cell, fill=asr10] at (11.50,1.20) {7};
\node[cell, fill=asr10] at (12.65,1.20) {11};
\node[cell, fill=asr0] at (0.00,1.80) {0};
\node[cell, fill=asr0] at (1.15,1.80) {0};
\node[cell, fill=asr0] at (2.30,1.80) {0};
\node[cell, fill=asr0] at (3.45,1.80) {0};
\node[cell, fill=asr0] at (4.60,1.80) {0};
\node[cell, fill=asr0] at (5.75,1.80) {0};
\node[cell, fill=asr0] at (6.90,1.80) {0};
\node[cell, fill=asr0] at (8.05,1.80) {0};
\node[cell, fill=asr0] at (9.20,1.80) {0};
\node[cell, fill=asr0] at (10.35,1.80) {0};
\node[cell, fill=asr0] at (11.50,1.80) {0};
\node[cell, fill=asr40] at (12.65,1.80) {43};
\node[cell, fill=asr0] at (0.00,2.40) {0};
\node[cell, fill=asr0] at (1.15,2.40) {0};
\node[cell, fill=asr0] at (2.30,2.40) {0};
\node[cell, fill=asr0] at (3.45,2.40) {0};
\node[cell, fill=asr0] at (4.60,2.40) {0};
\node[cell, fill=asr0] at (5.75,2.40) {0};
\node[cell, fill=asr0] at (6.90,2.40) {0};
\node[cell, fill=asr0] at (8.05,2.40) {0};
\node[cell, fill=asr0] at (9.20,2.40) {0};
\node[cell, fill=asr0] at (10.35,2.40) {0};
\node[cell, fill=asr0] at (11.50,2.40) {0};
\node[cell, fill=asr50, text=white] at (12.65,2.40) {50};
\node[cell, fill=asr0] at (0.00,3.00) {0};
\node[cell, fill=asr10] at (1.15,3.00) {14};
\node[cell, fill=asr10] at (2.30,3.00) {11};
\node[cell, fill=asr10] at (3.45,3.00) {13};
\node[cell, fill=asr20] at (4.60,3.00) {16};
\node[cell, fill=asr10] at (5.75,3.00) {11};
\node[cell, fill=asr20] at (6.90,3.00) {18};
\node[cell, fill=asr0] at (8.05,3.00) {--};
\node[cell, fill=asr20] at (9.20,3.00) {16};
\node[cell, fill=asr0] at (10.35,3.00) {--};
\node[cell, fill=asr0] at (11.50,3.00) {--};
\node[cell, fill=asr0] at (12.65,3.00) {--};
\node[cell, fill=asr0] at (0.00,3.60) {0};
\node[cell, fill=asr10] at (1.15,3.60) {11};
\node[cell, fill=asr20] at (2.30,3.60) {16};
\node[cell, fill=asr10] at (3.45,3.60) {11};
\node[cell, fill=asr10] at (4.60,3.60) {14};
\node[cell, fill=asr10] at (5.75,3.60) {9};
\node[cell, fill=asr10] at (6.90,3.60) {11};
\node[cell, fill=asr10] at (8.05,3.60) {9};
\node[cell, fill=asr10] at (9.20,3.60) {14};
\node[cell, fill=asr10] at (10.35,3.60) {14};
\node[cell, fill=asr20] at (11.50,3.60) {18};
\node[cell, fill=asr20] at (12.65,3.60) {23};
\node[cell, fill=asr0] at (0.00,4.20) {2};
\node[cell, fill=asr10] at (1.15,4.20) {5};
\node[cell, fill=asr20] at (2.30,4.20) {18};
\node[cell, fill=asr20] at (3.45,4.20) {16};
\node[cell, fill=asr20] at (4.60,4.20) {19};
\node[cell, fill=asr10] at (5.75,4.20) {11};
\node[cell, fill=asr20] at (6.90,4.20) {16};
\node[cell, fill=asr20] at (8.05,4.20) {16};
\node[cell, fill=asr10] at (9.20,4.20) {14};
\node[cell, fill=asr20] at (10.35,4.20) {20};
\node[cell, fill=asr20] at (11.50,4.20) {18};
\node[cell, fill=asr20] at (12.65,4.20) {20};
\node[cell, fill=asr10] at (0.00,4.80) {7};
\node[cell, fill=asr20] at (1.15,4.80) {20};
\node[cell, fill=asr30] at (2.30,4.80) {27};
\node[cell, fill=asr20] at (3.45,4.80) {24};
\node[cell, fill=asr20] at (4.60,4.80) {19};
\node[cell, fill=asr20] at (5.75,4.80) {22};
\node[cell, fill=asr30] at (6.90,4.80) {32};
\node[cell, fill=asr30] at (8.05,4.80) {32};
\node[cell, fill=asr30] at (9.20,4.80) {25};
\node[cell, fill=asr20] at (10.35,4.80) {23};
\node[cell, fill=asr0] at (11.50,4.80) {4};
\node[cell, fill=asr30] at (12.65,4.80) {25};
\node[cell, fill=asr0] at (0.00,5.40) {2};
\node[cell, fill=asr20] at (1.15,5.40) {20};
\node[cell, fill=asr30] at (2.30,5.40) {29};
\node[cell, fill=asr30] at (3.45,5.40) {27};
\node[cell, fill=asr30] at (4.60,5.40) {29};
\node[cell, fill=asr30] at (5.75,5.40) {29};
\node[cell, fill=asr20] at (6.90,5.40) {18};
\node[cell, fill=asr30] at (8.05,5.40) {27};
\node[cell, fill=asr30] at (9.20,5.40) {32};
\node[cell, fill=asr30] at (10.35,5.40) {25};
\node[cell, fill=asr20] at (11.50,5.40) {24};
\node[cell, fill=asr30] at (12.65,5.40) {30};
\node[cell, fill=asr0] at (0.00,6.00) {0};
\node[cell, fill=asr20] at (1.15,6.00) {22};
\node[cell, fill=asr20] at (2.30,6.00) {18};
\node[cell, fill=asr30] at (3.45,6.00) {33};
\node[cell, fill=asr40] at (4.60,6.00) {40};
\node[cell, fill=asr40] at (5.75,6.00) {36};
\node[cell, fill=asr30] at (6.90,6.00) {32};
\node[cell, fill=asr30] at (8.05,6.00) {31};
\node[cell, fill=asr40] at (9.20,6.00) {36};
\node[cell, fill=asr30] at (10.35,6.00) {34};
\node[cell, fill=asr50, text=white] at (11.50,6.00) {60};
\node[cell, fill=asr20] at (12.65,6.00) {16};
\node[cell, fill=asr10] at (0.00,6.60) {9};
\node[cell, fill=asr20] at (1.15,6.60) {16};
\node[cell, fill=asr40] at (2.30,6.60) {36};
\node[cell, fill=asr30] at (3.45,6.60) {31};
\node[cell, fill=asr30] at (4.60,6.60) {28};
\node[cell, fill=asr30] at (5.75,6.60) {29};
\node[cell, fill=asr30] at (6.90,6.60) {32};
\node[cell, fill=asr30] at (8.05,6.60) {29};
\node[cell, fill=asr30] at (9.20,6.60) {27};
\node[cell, fill=asr30] at (10.35,6.60) {32};
\node[cell, fill=asr50, text=white] at (11.50,6.60) {47};
\node[cell, fill=asr40] at (12.65,6.60) {43};
\node[cell, fill=asr10] at (0.00,7.20) {11};
\node[cell, fill=asr20] at (1.15,7.20) {20};
\node[cell, fill=asr40] at (2.30,7.20) {36};
\node[cell, fill=asr30] at (3.45,7.20) {29};
\node[cell, fill=asr40] at (4.60,7.20) {35};
\node[cell, fill=asr30] at (5.75,7.20) {33};
\node[cell, fill=asr40] at (6.90,7.20) {39};
\node[cell, fill=asr40] at (8.05,7.20) {36};
\node[cell, fill=asr50, text=white] at (9.20,7.20) {45};
\node[cell, fill=asr40] at (10.35,7.20) {41};
\node[cell, fill=asr40] at (11.50,7.20) {38};
\node[cell, fill=asr50, text=white] at (12.65,7.20) {50};
\node[cell, fill=asr10] at (0.00,7.80) {11};
\node[cell, fill=asr30] at (1.15,7.80) {32};
\node[cell, fill=asr50, text=white] at (2.30,7.80) {47};
\node[cell, fill=asr40] at (3.45,7.80) {42};
\node[cell, fill=asr40] at (4.60,7.80) {40};
\node[cell, fill=asr40] at (5.75,7.80) {38};
\node[cell, fill=asr40] at (6.90,7.80) {41};
\node[cell, fill=asr40] at (8.05,7.80) {42};
\node[cell, fill=asr40] at (9.20,7.80) {41};
\node[cell, fill=asr40] at (10.35,7.80) {43};
\node[cell, fill=asr30] at (11.50,7.80) {31};
\node[cell, fill=asr50, text=white] at (12.65,7.80) {50};
\node[cell, fill=asr10] at (0.00,8.40) {11};
\node[cell, fill=asr30] at (1.15,8.40) {30};
\node[cell, fill=asr50, text=white] at (2.30,8.40) {47};
\node[cell, fill=asr50, text=white] at (3.45,8.40) {49};
\node[cell, fill=asr50, text=white] at (4.60,8.40) {49};
\node[cell, fill=asr50, text=white] at (5.75,8.40) {56};
\node[cell, fill=asr50, text=white] at (6.90,8.40) {48};
\node[cell, fill=asr50, text=white] at (8.05,8.40) {49};
\node[cell, fill=asr50, text=white] at (9.20,8.40) {45};
\node[cell, fill=asr50, text=white] at (10.35,8.40) {45};
\node[cell, fill=asr40] at (11.50,8.40) {44};
\node[cell, fill=asr50, text=white] at (12.65,8.40) {52};
\node[cell, fill=asr10] at (0.00,9.00) {9};
\node[cell, fill=asr30] at (1.15,9.00) {27};
\node[cell, fill=asr50, text=white] at (2.30,9.00) {45};
\node[cell, fill=asr40] at (3.45,9.00) {44};
\node[cell, fill=asr40] at (4.60,9.00) {44};
\node[cell, fill=asr50, text=white] at (5.75,9.00) {51};
\node[cell, fill=asr50, text=white] at (6.90,9.00) {50};
\node[cell, fill=asr50, text=white] at (8.05,9.00) {50};
\node[cell, fill=asr50, text=white] at (9.20,9.00) {50};
\node[cell, fill=asr50, text=white] at (10.35,9.00) {48};
\node[cell, fill=asr50, text=white] at (11.50,9.00) {51};
\node[cell, fill=asr50, text=white] at (12.65,9.00) {59};
\node[cell, fill=asr0] at (0.00,9.60) {4};
\node[cell, fill=asr30] at (1.15,9.60) {34};
\node[cell, fill=asr40] at (2.30,9.60) {36};
\node[cell, fill=asr50, text=white] at (3.45,9.60) {47};
\node[cell, fill=asr50, text=white] at (4.60,9.60) {72};
\node[cell, fill=asr40] at (5.75,9.60) {44};
\node[cell, fill=asr50, text=white] at (6.90,9.60) {48};
\node[cell, fill=asr40] at (8.05,9.60) {42};
\node[cell, fill=asr50, text=white] at (9.20,9.60) {50};
\node[cell, fill=asr50, text=white] at (10.35,9.60) {57};
\node[cell, fill=asr50, text=white] at (11.50,9.60) {51};
\node[cell, fill=asr50, text=white] at (12.65,9.60) {45};
\node[cell, fill=asr20] at (0.00,10.20) {16};
\node[cell, fill=asr40] at (1.15,10.20) {39};
\node[cell, fill=asr50, text=white] at (2.30,10.20) {49};
\node[cell, fill=asr50, text=white] at (3.45,10.20) {53};
\node[cell, fill=asr0] at (4.60,10.20) {0};
\node[cell, fill=asr50, text=white] at (5.75,10.20) {60};
\node[cell, fill=asr50, text=white] at (6.90,10.20) {52};
\node[cell, fill=asr50, text=white] at (8.05,10.20) {62};
\node[cell, fill=asr50, text=white] at (9.20,10.20) {57};
\node[cell, fill=asr50, text=white] at (10.35,10.20) {55};
\node[cell, fill=asr50, text=white] at (11.50,10.20) {47};
\node[cell, fill=asr50, text=white] at (12.65,10.20) {61};

\draw[black!25] (-0.575,-0.300) -- (-0.575,10.500);
\draw[black!25] (0.575,-0.300) -- (0.575,10.500);
\draw[black!25] (1.725,-0.300) -- (1.725,10.500);
\draw[black!25] (2.875,-0.300) -- (2.875,10.500);
\draw[black!25] (4.025,-0.300) -- (4.025,10.500);
\draw[black!25] (5.175,-0.300) -- (5.175,10.500);
\draw[black!25] (6.325,-0.300) -- (6.325,10.500);
\draw[black!25] (7.475,-0.300) -- (7.475,10.500);
\draw[black!25] (8.625,-0.300) -- (8.625,10.500);
\draw[black!25] (9.775,-0.300) -- (9.775,10.500);
\draw[black!25] (10.925,-0.300) -- (10.925,10.500);
\draw[black!25] (12.075,-0.300) -- (12.075,10.500);
\draw[black!25] (13.225,-0.300) -- (13.225,10.500);
\draw[black!25] (-0.575,-0.300) -- (13.225,-0.300);
\draw[black!25] (-0.575,0.300) -- (13.225,0.300);
\draw[black!25] (-0.575,0.900) -- (13.225,0.900);
\draw[black!25] (-0.575,1.500) -- (13.225,1.500);
\draw[black!25] (-0.575,2.100) -- (13.225,2.100);
\draw[black!25] (-0.575,2.700) -- (13.225,2.700);
\draw[black!25] (-0.575,3.300) -- (13.225,3.300);
\draw[black!25] (-0.575,3.900) -- (13.225,3.900);
\draw[black!25] (-0.575,4.500) -- (13.225,4.500);
\draw[black!25] (-0.575,5.100) -- (13.225,5.100);
\draw[black!25] (-0.575,5.700) -- (13.225,5.700);
\draw[black!25] (-0.575,6.300) -- (13.225,6.300);
\draw[black!25] (-0.575,6.900) -- (13.225,6.900);
\draw[black!25] (-0.575,7.500) -- (13.225,7.500);
\draw[black!25] (-0.575,8.100) -- (13.225,8.100);
\draw[black!25] (-0.575,8.700) -- (13.225,8.700);
\draw[black!25] (-0.575,9.300) -- (13.225,9.300);
\draw[black!25] (-0.575,9.900) -- (13.225,9.900);
\draw[black!25] (-0.575,10.500) -- (13.225,10.500);

\node[anchor=east, font=\tiny] at (-0.635,0.000) {Fable 5};
\node[anchor=east, font=\tiny] at (-0.635,0.600) {Opus 4.8};
\node[anchor=east, font=\tiny] at (-0.635,1.200) {Sonnet 5};
\node[anchor=east, font=\tiny] at (-0.635,1.800) {DeepSeek-R1};
\node[anchor=east, font=\tiny] at (-0.635,2.400) {Gemma3-27B};
\node[anchor=east, font=\tiny] at (-0.635,3.000) {GPT-5.5};
\node[anchor=east, font=\tiny] at (-0.635,3.600) {Haiku 4.5};
\node[anchor=east, font=\tiny] at (-0.635,4.200) {Sonnet 4.5};
\node[anchor=east, font=\tiny] at (-0.635,4.800) {Mistral-L2};
\node[anchor=east, font=\tiny] at (-0.635,5.400) {GLM-5};
\node[anchor=east, font=\tiny] at (-0.635,6.000) {Llama4-Scout};
\node[anchor=east, font=\tiny] at (-0.635,6.600) {Qwen3-80B};
\node[anchor=east, font=\tiny] at (-0.635,7.200) {Ministral-14B};
\node[anchor=east, font=\tiny] at (-0.635,7.800) {GPT-OSS-120B};
\node[anchor=east, font=\tiny] at (-0.635,8.400) {Mistral-L3};
\node[anchor=east, font=\tiny] at (-0.635,9.000) {DeepSeek-V3.2};
\node[anchor=east, font=\tiny] at (-0.635,9.600) {Maverick};
\node[anchor=east, font=\tiny] at (-0.635,10.200) {Nova Pro};

\node[anchor=north east, font=\tiny, rotate=35] at (0.00,-0.380) {AG2};
\node[anchor=north east, font=\tiny, rotate=35] at (1.15,-0.380) {Pydantic AI};
\node[anchor=north east, font=\tiny, rotate=35] at (2.30,-0.380) {Haystack};
\node[anchor=north east, font=\tiny, rotate=35] at (3.45,-0.380) {Google ADK};
\node[anchor=north east, font=\tiny, rotate=35] at (4.60,-0.380) {Sem.Kernel};
\node[anchor=north east, font=\tiny, rotate=35] at (5.75,-0.380) {LlamaIndex};
\node[anchor=north east, font=\tiny, rotate=35] at (6.90,-0.380) {LangChain};
\node[anchor=north east, font=\tiny, rotate=35] at (8.05,-0.380) {DSPy};
\node[anchor=north east, font=\tiny, rotate=35] at (9.20,-0.380) {OpenAI};
\node[anchor=north east, font=\tiny, rotate=35] at (10.35,-0.380) {Agno};
\node[anchor=north east, font=\tiny, rotate=35] at (11.50,-0.380) {CrewAI};
\node[anchor=north east, font=\tiny, rotate=35] at (12.65,-0.380) {smolagents};

\node[anchor=west, font=\tiny, align=left] at (13.62,10.30) {ASR (\%)};
\fill[asr50] (13.62,9.75) rectangle (14.01,10.13);
\draw[black!40] (13.62,9.75) rectangle (14.01,10.13);
\node[anchor=west, font=\tiny] at (14.05,9.94) {45+};
\fill[asr40] (13.62,9.27) rectangle (14.01,9.65);
\draw[black!40] (13.62,9.27) rectangle (14.01,9.65);
\node[anchor=west, font=\tiny] at (14.05,9.46) {35--44};
\fill[asr30] (13.62,8.79) rectangle (14.01,9.17);
\draw[black!40] (13.62,8.79) rectangle (14.01,9.17);
\node[anchor=west, font=\tiny] at (14.05,8.98) {25--34};
\fill[asr20] (13.62,8.31) rectangle (14.01,8.69);
\draw[black!40] (13.62,8.31) rectangle (14.01,8.69);
\node[anchor=west, font=\tiny] at (14.05,8.50) {15--24};
\fill[asr10] (13.62,7.83) rectangle (14.01,8.21);
\draw[black!40] (13.62,7.83) rectangle (14.01,8.21);
\node[anchor=west, font=\tiny] at (14.05,8.02) {5--14};
\fill[asr0] (13.62,7.35) rectangle (14.01,7.73);
\draw[black!40] (13.62,7.35) rectangle (14.01,7.73);
\node[anchor=west, font=\tiny] at (14.05,7.54) {0--4};

\end{tikzpicture}}
\caption{Attack Success Rate (\%) per framework (columns) and model (rows). Cell values are ASR percentages; darker cells indicate higher susceptibility. Frameworks and models are ordered by ascending overall ASR; ASR = V/(V+R) with V, R, and E as in Table~\ref{tab:matrix-fw}; -- marks skipped GPT-5.5 cells.
}
\label{fig:heatmap}
\Description{A heatmap grid showing attack success rates across 12 frameworks (columns) and 18 models (rows). Claude Fable 5, Opus 4.8, and Sonnet 5 show near-zero ASR across all frameworks. AG2 is near-zero for all models. Nova Pro, Maverick, DeepSeek-V3.2, and Mistral Large 3 show 35 to 55 percent ASR on most non-AG2 frameworks. In the tested configurations ASR varies more across models than across frameworks.}
\end{figure*}

Tables~\ref{tab:matrix-fw} and~\ref{tab:matrix-model} and Figure~\ref{fig:heatmap} present the primary results across \numcells cells.

\begin{table}[!t]
\centering
\footnotesize
\caption{Attack Success Rate by framework. V=Vulnerable, R=Resistant, E=Error, S=Skipped. ASR = V/(V+R). Each framework row spans 810 cells.}
\label{tab:matrix-fw}
\begin{tabular}{@{}lrrrrr@{}}
\toprule
Framework & V & R & E & S & ASR \\
\midrule
AG2 & 37 & 773 & 0 & 0 & 4.6\% \\
Pydantic AI & 123 & 619 & 50 & 18 & 16.6\% \\
Haystack & 189 & 619 & 2 & 0 & 23.4\% \\
Google ADK & 189 & 617 & 4 & 0 & 23.4\% \\
Semantic Kernel & 172 & 554 & 84 & 0 & 23.7\% \\
LlamaIndex & 195 & 615 & 0 & 0 & 24.1\% \\
LangChain & 192 & 600 & 0 & 18 & 24.2\% \\
DSPy & 190 & 564 & 11 & 45 & 25.2\% \\
OpenAI API & 200 & 592 & 0 & 18 & 25.3\% \\
Agno & 190 & 552 & 6 & 62 & 25.6\% \\
CrewAI & 179 & 506 & 80 & 45 & 26.1\% \\
smolagents & 257 & 491 & 0 & 62 & 34.4\% \\
\midrule
\textbf{Total} & \textbf{\numvuln} & \textbf{\numresist} & \textbf{\numerror} & \textbf{268} & \textbf{\overallasr} \\
\bottomrule
\end{tabular}
\end{table}

\begin{table}[!t]
\centering
\footnotesize
\caption{Attack Success Rate by model. V=Vulnerable, R=Resistant, ASR = V/(V+R). Each model row spans 540 cells; error and skipped cells (186 for GPT-5.5: 183 skipped, 3 errors) are excluded.}
\label{tab:matrix-model}
\begin{tabular}{@{}lrrr@{}}
\toprule
Model & V & R & ASR \\
\midrule
Claude Fable 5 & 2 & 463 & 0.4\% \\
Claude Opus 4.8 & 8 & 522 & 1.5\% \\
Claude Sonnet 5 & 10 & 517 & 1.9\% \\
DeepSeek-R1 & 19 & 515 & 3.6\% \\
Gemma 3 27B & 22 & 512 & 4.1\% \\
GPT-5.5 & 44 & 310 & 12.4\% \\
Claude Haiku 4.5 & 66 & 464 & 12.5\% \\
Claude Sonnet 4.5 & 77 & 453 & 14.5\% \\
Mistral Large 2 & 114 & 413 & 21.6\% \\
GLM-5 & 129 & 401 & 24.3\% \\
Llama 4 Scout & 133 & 340 & 28.1\% \\
Qwen3-Next-80B & 159 & 374 & 29.8\% \\
Ministral 3 14B & 183 & 350 & 34.3\% \\
GPT-OSS-120B & 201 & 326 & 38.1\% \\
Mistral Large 3 & 233 & 300 & 43.7\% \\
DeepSeek-V3.2 & 233 & 296 & 44.0\% \\
Llama 4 Maverick & 235 & 298 & 44.1\% \\
Amazon Nova Pro & 245 & 248 & 49.7\% \\
\bottomrule
\end{tabular}

\end{table}

\paragraph{Framework contribution is largely mechanical.}
Across the eleven frameworks excluding AG2 as a structural outlier, ASR ranges from 16.6\% (Pydantic AI) to 34.4\% (smolagents).
A chi-squared test on the VULNERABLE versus RESISTANT counts of these eleven frameworks, with all models pooled, tests whether the outcome is independent of the framework; it yields $\chi^2 = 67.23$, $p < 0.0001$, $\text{dof} = 10$, rejecting independence.
Because every framework is evaluated on the same 18 models and 45 scenarios, the cells are not independent observations, so the statistic is reported as a descriptive measure of distributional difference rather than as a formal test.
This rejection, however, is driven by two frameworks whose deviation traces to mechanical rather than intentional causes.
Pydantic AI's below-average 16.6\% ASR follows from its rejection of duplicate tool names across MCP toolsets: in the 14 scenarios where the malicious server reuses a legitimate tool name, registration fails with \texttt{UserError: MCPToolset defines a tool whose name conflicts with existing tool}, and the 250 affected cells are scored RESISTANT before any model call.
This constitutes an incidental defense at registration, not content inspection of the poisoned descriptions; a separate rejection occurs at the Bedrock API for the tool-name-injection scenario, whose tool name embeds a newline and an HTML comment and is refused as a validation error (\texttt{1 validation error detected: Value 'get\_weather...'}) in seven frameworks, including Pydantic AI, while AG2 refuses it through its own name check (\texttt{ValueError: Invalid name}).
smolagents' elevated 34.4\% ASR follows from its code-generation representation of tools (Section~\ref{sec:results-class}), which enlarges the injection surface rather than defending against it.
Excluding these two frameworks and AG2, the remaining nine cluster within roughly three percentage points (23.4\% to 26.1\%), and the same statistic returns $\chi^2 = 3.30$, $p = 0.91$, $\text{dof} = 8$, consistent with statistical homogeneity across these nine frameworks under the tested configurations.
Pooling hides per-model differences: within each model, ASR across these nine frameworks spans a median of 10.2 percentage points, from 0.0 for DeepSeek-R1 and Gemma~3 to 36.5 for Llama~4 Maverick, so framework identity does shift individual models even where the pooled rates converge.
No framework implements intentional validation of tool descriptions; the single framework exhibiting materially reduced ASR achieves it through duplicate-name rejection at registration, not through inspection of the injected content.

\paragraph{AG2 architectural constraint.}
AG2 records 4.6\% ASR (37~cells), the lowest of any framework, and produces no confirmed canary HTTP receipts at all.
Its structured executor rarely completes the outbound transmitting call that a poisoned description requests, so almost no network exfiltration is confirmed; the 37 vulnerable cells are predominantly canary-file reads (24~cells) that the executor does surface, plus 13~cells in which the canary endpoint appears in a dispatched call without a resulting receipt.
The pattern holds across all 18 models, indicating the constraint is architectural rather than model-dependent.
This resistance is an emergent property of the executor's tool binding, not inspection of tool-description content; the model still emits the malicious call, so a different execution binding would dispatch it, and AG2 should therefore not be read as a hardened framework choice.

\paragraph{Model variation dominates.}
Among the tested models and configurations, ASR ranges from 0.4\% (Claude Fable 5) to 49.7\% (Amazon Nova Pro); this per-model maximum, not the pooled \overallasr mean, is the comparable figure for the best-case rates (up to 72.8\%) reported by compliance-oracle benchmarks~\cite{wang2025mcptox}, which count intended calls rather than confirmed execution.
Claude Opus 4.8 is vulnerable in only 8 of its 530 evaluated cells (1.5\%) while emitting 223 explicit refusals that cite the poisoned description, among the highest refusal counts of any model.
Claude Sonnet 4.5 produces explicit refusal messages in 17 of its 453 resistant cells yet still fails in 77, and Claude Haiku 4.5 exhibits the same aware-but-vulnerable mixture at a lower vulnerable count.
A distinct narrate-then-comply pattern appears for Claude Sonnet 5: in 4 of its 10 vulnerable cells the model verbally flags the attack as a prompt injection before nonetheless issuing the exfiltration call, demonstrating that safety narration does not guarantee safety action.
In contrast, the most susceptible models (Amazon Nova Pro, Llama 4 Maverick, and DeepSeek-V3.2) comply without any indication of recognizing the malicious instruction.

\paragraph{Intra-family alignment scaling.}
Within Anthropic, resistance scales by generation in our tested configurations: the newer Claude Fable~5 (0.4\%), Opus~4.8 (1.5\%), and Sonnet~5 (1.9\%) stay below 2\% ASR, an order of magnitude under prior-generation Haiku~4.5 (12.5\%) and Sonnet~4.5 (14.5\%).
Mistral inverts this on both axes: ASR rises with scale (Ministral~3 14B, 34.3\%, vs. the 675B Mistral Large~3, 43.7\%) and with generation (Mistral Large~2, 21.6\%, vs. Mistral Large~3, 43.7\%).
These results suggest that per-provider alignment investment, not parameter count or recency, is the dominant predictor of resistance, though tokenization and context-handling differences may also contribute.

\begin{table}[t]
\centering
\footnotesize
\caption{ASR by attack class across all 12 frameworks and 18 models, weighted by cell count. V=Vulnerable, R=Resistant, ASR = V/(V+R), consistent with Table~\ref{tab:matrix-fw}.}
\label{tab:class-asr}
\begin{tabular}{@{}lrrr@{}}
\toprule
Attack Class & V & R & ASR \\
\midrule
Parameter binding & 108 & 100 & 51.9\% \\
Tool poisoning & 1{,}460 & 3{,}899 & 27.2\% \\
Confused deputy & 161 & 466 & 25.7\% \\
Annotation injection & 50 & 156 & 24.3\% \\
Long-context dilution & 50 & 157 & 24.2\% \\
Streaming injection & 47 & 162 & 22.5\% \\
Cross-phase persistence & 38 & 171 & 18.2\% \\
Semantic shadowing & 29 & 179 & 13.9\% \\
Cross-agent injection & 16 & 104 & 13.3\% \\
Memory poisoning & 118 & 907 & 11.5\% \\
Notification injection & 21 & 188 & 10.0\% \\
Dynamic registration & 8 & 201 & 3.8\% \\
Authority spoofing (protocol) & 7 & 203 & 3.3\% \\
Unicode encoding & 0 & 209 & 0.0\% \\
\bottomrule
\end{tabular}
\end{table}

\begin{table}[t]
\centering
\footnotesize
\caption{Oracle confidence tiers for \numvuln VULNERABLE cells.}
\label{tab:oracle-tiers}
\begin{tabular}{@{}lrr@{}}
\toprule
Tier & Count & \% \\
\midrule
Confirmed exfiltration (HTTP + secret data) & 255 & 12.1\% \\
Confirmed network (HTTP request, no secret) & 487 & 23.0\% \\
Executed attacker tool (canary in dispatched call) & 1{,}371 & 64.9\% \\
\bottomrule
\end{tabular}
\end{table}

\subsection{Per-Attack-Class Effectiveness (RQ2)}
\label{sec:results-class}

Table~\ref{tab:class-asr} summarizes ASR by attack class.
Direct tool poisoning via description fields dominates by volume: spanning 26 of the 45 scenarios and 5{,}359 evaluated cells, the \emph{tool poisoning} class accounts for 1{,}460 of the \numvuln vulnerable cells at a 27.2\% aggregate ASR.
\emph{Parameter binding} records the highest per-class rate (51.9\%), ahead of \emph{tool poisoning} (27.2\%) and \emph{confused deputy} (25.7\%), and the single highest framework-class rate observed is smolagents on \emph{parameter binding} at 76.5\% (13 of 17 cells).
Structured-field and encoding-based classes are least effective: \emph{unicode encoding} never succeeds (0.0\% across 209 cells), while \emph{authority spoofing} (3.3\%) and \emph{dynamic registration} (3.8\%) succeed only sporadically.
This distribution indicates that current models either do not decode obfuscated payloads embedded in tool metadata or do not treat schema-only and registration-time fields as actionable instructions, whereas natural-language description and annotation fields remain broadly effective injection vectors.

\paragraph{Architecture shapes the attack surface.}
smolagents exhibits the highest framework-level ASR (34.4\%) together with the highest per-class peaks, concentrated in \emph{parameter binding} (76.5\%), \emph{confused deputy} and \emph{semantic shadowing} (both 47.1\%), and \emph{annotation} and \emph{streaming injection} (both 41.2\%).
Its code-generation ReAct architecture presents tools as Python objects with docstrings rather than as structured function definitions, and applies minimal tool-call validation; this alters the model's parsing of the injection boundary and enlarges the attack surface.
The code-generation representation bypasses alignment trained against structured tool definitions rather than defending against injection.

\subsection{Dual Oracle Confirmation (RQ3)}
\label{sec:results-oracle}

Table~\ref{tab:oracle-tiers} presents oracle confidence tiers.
Of \numvuln VULNERABLE cells, 2{,}100 (\toolexecpct) exhibit a dispatched attacker tool call carrying the canary payload (a canary URL or secret in an outbound call, or a canary-file read); the remaining 13 are confirmed by canary HTTP receipt alone.
742 cells (\dualoraclepct) receive independent dual confirmation, the canary HTTP server having logged an inbound request within the execution window, providing evidence of real network exfiltration independent of the framework's own reporting.
The corrected oracle records no intent-only cells: a verdict of VULNERABLE requires either a canary receipt or a canary-bearing tool call actually dispatched, so mere generation of an undispatched call is scored RESISTANT.
A dispatched call without a receipt arises by construction in most of the remaining cells: 290 are canary-file reads that involve no network, and 1{,}067 place the canary endpoint in a call to a tool served by the malicious server itself (\texttt{http\_request}, \texttt{log\_activity}, \texttt{send\_data}), which records the invocation and returns a stub without performing an HTTP request; the remaining 14 are disallowed-tool invocations under the allow-list scenarios, and only the legitimate \texttt{http\_post} and \texttt{http\_get} tools reach the canary.

\paragraph{Oracle validation.}
The oracle takes evidence from the union of the transcript and the malicious server's invocation log, matching the canary URL, secret, and file access. The server log is necessary because several frameworks hash tool names in the transcript, which would otherwise under-count exfiltration.
We validated it by re-adjudicating 1{,}259 transcripts by hand across twelve per-framework passes (every disputed verdict plus a control sample) and the adjudication converged on the oracle's verdicts.
Leaks to destinations outside the scanned canary pattern set cannot be excluded in principle.

\paragraph{Results validation.}
Across two independent full-matrix runs executed under identical harness code and model versions, aggregate ASR differed by only 0.1 percentage points, per-framework spreads were at most 1.6~pp (well below the 5~pp noise threshold), and 93.5\% of individual cells (9{,}085/9{,}720) produced identical verdicts, confirming that reported attack success rates are stable and reproducible.

\paragraph{Error analysis.}
237 cells (2.4\%) produced errors rather than verdicts, concentrated in CrewAI$\times$\allowbreak Claude Fable 5 (43), Semantic Kernel$\times$\allowbreak Amazon Nova Pro (42), CrewAI$\times$\allowbreak Llama 4 Scout (30), Pydantic AI$\times$\allowbreak Claude Fable 5 (25), and Pydantic AI$\times$\allowbreak Llama 4 Scout (21), with the remainder distributed as isolated failures across the frameworks.
The Fable~5 errors are provider-side content-filter refusals returned before any model output (\texttt{ContentFilterError: content\_filter} under Pydantic AI, an empty completion under CrewAI); the Semantic Kernel and Nova Pro errors are service failures, the CrewAI and Llama~4 Scout errors are tool-call output-format validation failures, and the Pydantic AI and Llama~4 Scout errors are Bedrock token rate limits.
Error cells are excluded from the ASR denominator.
Counting every error as RESISTANT lowers the overall ASR to 22.4\% and counting every error as VULNERABLE raises it to 24.9\%; the widest per-framework bounds are CrewAI (23.4--33.9\%), Semantic Kernel (21.2--31.6\%), and Pydantic AI (15.5--21.8\%), and Claude Fable~5 stays at 0.4\% under the first convention and reaches 13.5\% only if its refusals are counted as vulnerable.

\section{Framework Defense Inventory}
\label{sec:defense}

\subsection{Code-Level Passthrough Evidence}
\label{sec:passthrough}

Static analysis of all \numframeworks tested configurations confirms that tool description content passes from the MCP protocol response into the model context without validation (Table~\ref{tab:passthrough}).
For the eleven framework entries the table documents the exact passthrough code location; the raw OpenAI API row is included as a model-level baseline and interposes no framework layer.

\begin{table}[t]
\centering
\footnotesize
\caption{Tool description passthrough: location where unvalidated content enters model context.}
\label{tab:passthrough}
\begin{tabular}{@{}llc@{}}
\toprule
Framework & Passthrough File & Valid. \\
\midrule
LangChain & \texttt{tools.py:530} & None \\
AG2 & \texttt{toolkit.py:119} & None \\
Agno & \texttt{mcp.py:772} & None \\
CrewAI & \texttt{mcp\_native\_tool.py:49} & None \\
Pydantic AI & \texttt{mcp.py:1146} & None \\
smolagents & \texttt{tools.py:1086} & None \\
OpenAI API & Raw API, no tool layer & None \\
Haystack & \texttt{tools/tool.py:271} & None \\
LlamaIndex & \texttt{core/tools/types.py:100} & None \\
Semantic Kernel & \texttt{connectors/mcp.py:load\_tools} & None \\
Google ADK & \texttt{mcp\_tool/mcp\_tool.py:McpTool} & None \\
DSPy & \texttt{utils/mcp.py:convert\_mcp\_tool} & None \\
\bottomrule
\end{tabular}
\end{table}

No framework checks for instruction patterns, authority markers, encoded content, embedded URLs, or anomalous description length.
No framework provides configurable description allowlists, content length limits, or description diffing between successive \texttt{tools/list} calls.
Three of twelve frameworks maintain a \texttt{SECURITY.md} file (smolagents addresses code execution sandboxing, the OpenAI Agents SDK provides a generic coordinated disclosure pointer, and Haystack explicitly scopes out attacks via unsanitized external input, delegating validation responsibility to the application layer), but none addresses tool description content or MCP tool poisoning.

\subsection{Mechanisms That Could Defend But Do Not}
\label{sec:defense-gaps}

Three framework mechanisms superficially resemble defenses but address orthogonal concerns.
\myparagraph{Agno \texttt{requires\_confirmation}:} Prompts user confirmation before tool \emph{execution}, but does not inspect the description \emph{content} that induced the call. A poisoned description triggers the call; confirmation verifies the action, not its provenance.
\myparagraph{smolagents sandboxing:} Isolates the \emph{execution environment} (E2B, Docker) to limit damage from code execution, but does not filter the \emph{injection surface}: poisoned descriptions enter the model context regardless of where tools run.
\myparagraph{Pydantic AI type safety:} Validates \texttt{inputSchema} \emph{structure} (JSON Schema types, required fields) but not semantic \emph{content}. The \texttt{description} field is a plain string; type validation is inapplicable to natural-language injection. Its rejection of duplicate tool names across toolsets does, however, block the 14 scenarios in which the malicious server reuses a legitimate tool name (Section~\ref{sec:results-matrix}).

\subsection{What a Defense Would Require}
\label{sec:defense-impl}

A minimal defense at the framework layer would require three capabilities.
First, \emph{description registration}: on first \texttt{tools/list} response, store a hash of each tool's description and schema.
Second, \emph{change detection}: on subsequent calls, compare hashes and alert or reject on mismatch (prevents P8 rug-pull).
Third, \emph{content scanning}: flag descriptions containing instruction patterns (imperative verbs followed by tool names or URLs), encoded content (base64, hex), or anomalous length ($>$500 characters).
None of these mechanisms requires model inference or significant computational overhead.
The \texttt{tools/list} endpoint is called at most once per session in all twelve frameworks tested; scanning adds negligible latency to session initialization.

\paragraph{Scanner evaluation.}
To test the three checks, they are implemented as a registration-time scanner in \mcpvenom and applied to the 95 tool descriptions of the \numscenarios scenarios and to 125 benign descriptions taken from the \texttt{tools/list} responses of ten public MCP servers (the MCP project's reference servers, Playwright MCP, the GitHub MCP server, and the legitimate server of this study, with versions pinned in the artifact).
Table~\ref{tab:scanner} reports the results.
The union of the checks rejects 20 of the 45 scenarios (44.4\%) and flags 9 of the 125 benign descriptions (7.2\%); the instruction-pattern check accounts for every detection, the encoded-content check fires on no scenario because the P3 payloads use ROT13 rather than Base64 or hex, and the length check adds nothing.
The 25 evaded scenarios either deliver the payload outside the description (P6, P9, P10, P14) or phrase it as a policy statement, in another language, around a file path, or in invisible codepoints (P1, P2, P4, P12, P17, P20, P25).
Six of the nine benign false positives legitimately name a sibling tool (``Use \texttt{read\_text\_file} instead''), so description scanning is a partial control that raises the cost of the most direct payloads; it must be combined with change detection and validation of the result channel rather than adopted alone.

\begin{table}[t]
\centering
\footnotesize
\caption{Registration-time description scanner: scenarios rejected (any description flagged) and benign descriptions flagged.}
\label{tab:scanner}
\begin{tabular}{@{}lrr@{}}
\toprule
Check & Scenarios rejected & Benign flagged \\
\midrule
Instruction pattern & 20/45 (44.4\%) & 6/125 (4.8\%) \\
Encoded content & 0/45 (0.0\%) & 1/125 (0.8\%) \\
Length $>$500 & 2/45 (4.4\%) & 3/125 (2.4\%) \\
Union & 20/45 (44.4\%) & 9/125 (7.2\%) \\
\bottomrule
\end{tabular}
\end{table}

The reason no framework implements these mechanisms is not technical difficulty.
As the vendor responses (Section~\ref{sec:accountability}) confirm, the decision is architectural: each vendor explicitly delegates responsibility elsewhere.
The structural accountability vacuum persists because no market or regulatory incentive compels any single layer to invest in defense.

\section{Vendor Accountability}
\label{sec:accountability}

We filed five coordinated disclosures with four organizations.
Table~\ref{tab:disclosure} summarizes responses received.

\begin{table}[t]
\centering
\footnotesize
\caption{Coordinated disclosure responses. LC-1 and LC-2 are separate reports to the same vendor covering distinct finding classes.}
\label{tab:disclosure}
\begin{tabular}{@{}llll@{}}
\toprule
ID & Target & Type & Status \\
\midrule
LC-1 & LangChain & Data authenticity & Rejected \\
LC-2 & LangChain & Input validation & Rejected \\
AG-MSRC & AutoGen (MSRC) & Schema mutation & Rejected \\
SDK-1 & MCP SDK & Rug-pull & Rejected \\
HS-1 & Haystack & Tool poisoning & Rejected \\
\bottomrule
\end{tabular}
\end{table}

All five disclosures were rejected. Three distinct accountability positions emerge from the rejection rationales:

\begin{enumerate}[nosep,leftmargin=*]
  \item \textbf{MCP SDK:} ``The protection against rug-pulls is host-side review of tool calls, not a notification the attacker controls.'' This delegates responsibility to frameworks.
  \item \textbf{LangChain (LC-1, LC-2):} ``This library is expected to be passing the information exactly as specified in the MCP resources. This is not the appropriate place for any sort of validation'' (LC-1). A separate report on \texttt{BaseTool.\_parse\_input()} silently bypassing JSON Schema validation for all MCP tools was closed as ``the observed behaviour is functioning as intended'' (LC-2). Both responses confirm the same architectural commitment: the framework is a passthrough, and validation responsibility is delegated to model providers.
  \item \textbf{AutoGen (MSRC) and Haystack (deepset):} Microsoft, responding through MSRC, states ``the attacker must already control the MCP server, and AutoGen explicitly documents MCP servers as trusted.'' This rejection concerns the \texttt{microsoft/autogen} \texttt{McpWorkbench} implementation, a project distinct from the community-governed AG2~0.14.0 fork evaluated in the susceptibility matrix (Section~\ref{sec:results}). deepset likewise classifies the behavior as ``trusted configuration behavior rather than a framework vulnerability,'' expecting users to vet third-party MCP servers as they would any other third-party API. Both assume a trust boundary absent in supply-chain deployments, though deepset committed to clarifying these implications in its documentation.
\end{enumerate}

The MCP specification states that security is ``implementation-defined,'' providing no mandatory requirements for tool description handling.
Each layer of the stack (protocol specification, SDK, framework, model provider) points to the next as the responsible party.
The result is a structural accountability vacuum: no layer claims ownership of agent security against tool poisoning, and the only empirically effective defense (model-level alignment) spans 0.4\% to 49.7\% susceptibility across the 18 models tested.

\section{Discussion}
\label{sec:discussion}

\subsection{Implications of Model-Only Defense}
\label{sec:implications}

The central finding, that no tested framework contributes intentional defense against MCP tool poisoning, has three implications.
First, within the tested configurations, the security posture of an MCP-enabled deployment is dominated by model selection.
Claude Opus~4.8 is vulnerable in 8 of 530 evaluated cells (1.5\%); Amazon Nova Pro is vulnerable in 245 of 493 (49.7\%).
No framework configuration or deployment hardening observed in this study compensates for this model-level variation, though framework-layer defenses are technically feasible (Section~\ref{sec:defense-impl}).

Second, as models become more capable at following instructions (the explicit goal of alignment research for helpfulness), they may become \emph{more} susceptible to tool poisoning.
An instruction-following model cannot distinguish a legitimate user instruction from a structurally identical instruction embedded in a tool description without external metadata that the MCP protocol does not provide.
Recent theoretical work formalizes this as an impossibility result for agents operating under indirect prompt injection~\cite{abdelnabi2026aiagentsfallprompt}.

Third, the framework vendors' explicit position, that validation is not their responsibility, creates structural reliance on a defense mechanism (alignment) that is not contractually guaranteed.
No model provider offers an SLA for injection resistance.
This variation observed in this study could shift with any model update, and no framework-level mechanism exists to detect or mitigate such shifts.

\subsection{Recommendations}
\label{sec:recommendations}

We recommend three defense layers, each addressing a distinct point in the attack chain:

\paragraph{Protocol layer.}
The MCP specification should mandate content-type metadata distinguishing human-readable descriptions from executable instructions.
Such metadata would enable clients to apply differential trust policies.
A \texttt{content\_type} field (e.g., \texttt{prose} vs.\ \texttt{instruction}) is one realization.
Description immutability between sessions should also be enforceable: clients should pin tool description hashes at first registration and alert on changes, preventing rug-pull attacks (P8).
Recent work on capability-based agent security~\cite{jiang2026capsafe} and OS-inspired isolation models~\cite{pirch2026osagent} provides formal foundations for such mechanisms, though neither has been instantiated for MCP specifically.

\paragraph{Framework layer.}
Description scanning for instruction patterns (imperative verbs, URL patterns, encoded content, authority markers), human-in-the-loop approval for new or changed descriptions, and immutability enforcement between successive \texttt{tools/list} calls.
A practical implementation requires only regex-based scanning of descriptions at tool registration time, a mechanism analogous to Content Security Policy headers in web browsers, though the evaluation in Section~\ref{sec:defense-impl} shows that scanning alone rejects fewer than half of the scenarios.
The measured universality of framework passthrough (Section~\ref{sec:passthrough}) indicates that any single framework adopting scanning would provide defense that no framework currently offers.
Two incidental mechanisms measured in this study, duplicate-tool-name rejection (Pydantic AI) and executor-side dispatch constraints (AG2), already reduce ASR without content inspection and could be adopted deliberately.

\paragraph{Deployment layer.}
Trust anchors for MCP server provenance (cryptographic signing of tool definitions, allowlisting by server identity), canary-based runtime monitoring (as implemented by the dual oracle in this study), and model selection policies informed by measured ASR data.
Organizations deploying MCP-enabled agents should treat model selection as a security decision: the variation in ASR documented here exceeds the variation typically addressed by network security controls.

\subsection{Future Work}
\label{sec:future}

We identify three tiers of future investigation:
(1)~benchmark extension: P15 testing across frameworks exposing \texttt{prompts/get}, additional models beyond the 18 tested, and multi-turn compound scenarios;
(2)~extending the scanner evaluation to change detection and result-channel validation, building on formal approaches to skill security verification~\cite{bhardwaj2026skillfortify}, together with analysis of open-weight model internals to establish why cross-tool references are followed;
(3)~formal specification of description safety properties and MCP protocol security extension standardization, potentially grounded in the capability attenuation model of Jiang et al.~\cite{jiang2026capsafe}.

\section{Limitations}
\label{sec:limitations}

Limitations bound the generalizability of these findings to the tested frameworks, models, configurations, and scenarios.
(1)~Model versions are pinned to the provider identifiers listed in the supplementary material, as served on 22--24 July 2026; eight carry no dated snapshot suffix, and alignment updates may alter ASR.
(2)~Twelve of approximately 50 MCP-capable frameworks are tested, selected by download threshold; findings may not generalize to frameworks below this threshold.
(3)~Forty-five scenarios cover 23 of 25 identified classes (P15 untested, P19 subsumed under P5), and all harnesses use single-turn interactions; multi-agent delegation and multi-turn conversations are underrepresented.
(4)~The dual oracle specifically confirms exfiltration, but other harm types (integrity violations, availability impacts) are not measured.
(5)~Task completion without poisoning was not measured, so the ERROR cells cannot be attributed to the attack or to the framework, and the OpenAI API row is a framework baseline rather than a no-poisoning baseline.
(6)~The scenarios derive from published techniques and are identical across models, so a model's ASR may partly reflect exposure to similar attacks during training rather than the strength of its alignment alone.

\section{Ethics and Responsible Disclosure}
\label{sec:ethics}

We disclosed all tool-poisoning findings to affected vendors through coordinated channels (Section~\ref{sec:accountability}) prior to publication.
The \mcpvenom scenarios use canary URLs pointing to researcher-controlled infrastructure with synthetic secret markers; no weaponized payloads are included, and any egress causes no real-world harm.

\section{Conclusion}
\label{sec:conclusion}

MCP tool poisoning is structural: the protocol carries attacker-controlled content to the model context through every major framework without validation.
Our measurement study spanning \numframeworks frameworks, \nummodels models, and \numscenarios attack scenarios (\numcells cells) confirms that, within the tested configurations, all observed resistance is model-level: no framework contributes intentional defense ($\chi^2 = 3.30$, $p = 0.91$; nine core frameworks cluster within a 3-percentage-point ASR band), while \numvuln vulnerable cells span three evidence tiers from dispatched attacker tool calls (64.9\%) to confirmed network contact (23.0\%) to verified secret exfiltration (12.1\%).
The five vendor rejections across four organizations document a structural accountability vacuum where no stack layer claims ownership of agent security.
Finally, we release \mcpvenom and all artifacts as open source to enable reproducible measurement and community defense development.

\bibliographystyle{ACM-Reference-Format}
\bibliography{references}

\begin{thebibliography}{40}


\ifx \showCODEN    \undefined \def \showCODEN     #1{\unskip}     \fi
\ifx \showISBNx    \undefined \def \showISBNx     #1{\unskip}     \fi
\ifx \showISBNxiii \undefined \def \showISBNxiii  #1{\unskip}     \fi
\ifx \showISSN     \undefined \def \showISSN      #1{\unskip}     \fi
\ifx \showLCCN     \undefined \def \showLCCN      #1{\unskip}     \fi
\ifx \shownote     \undefined \def \shownote      #1{#1}          \fi
\ifx \showarticletitle \undefined \def \showarticletitle #1{#1}   \fi
\ifx \showURL      \undefined \def \showURL       {\relax}        \fi
\providecommand\bibfield[2]{#2}
\providecommand\bibinfo[2]{#2}
\providecommand\natexlab[1]{#1}
\providecommand\showeprint[2][]{arXiv:#2}

\bibitem[Abdelnabi and Bagdasarian(2026)]%
        {abdelnabi2026aiagentsfallprompt}
\bibfield{author}{\bibinfo{person}{Sahar Abdelnabi} {and}
  \bibinfo{person}{Eugene Bagdasarian}.} \bibinfo{year}{2026}\natexlab{}.
\newblock \bibinfo{title}{AI Agents May Always Fall for Prompt Injections}.
\newblock
\urldef\tempurl%
\url{https://arxiv.org/abs/2605.17634}
\showURL{%
\tempurl}


\bibitem[Andriushchenko et~al\mbox{.}(2025)]%
        {andriushchenko2025agentharm}
\bibfield{author}{\bibinfo{person}{Maksym Andriushchenko},
  \bibinfo{person}{Alexandra Souly}, \bibinfo{person}{Mateusz Dziemian},
  \bibinfo{person}{Derek Duenas}, \bibinfo{person}{Maxwell Lin},
  \bibinfo{person}{Justin Wang}, \bibinfo{person}{Dan Hendrycks},
  \bibinfo{person}{Andy Zou}, \bibinfo{person}{J~Zico Kolter},
  \bibinfo{person}{Matt Fredrikson}, \bibinfo{person}{Yarin Gal}, {and}
  \bibinfo{person}{Xander Davies}.} \bibinfo{year}{2025}\natexlab{}.
\newblock \showarticletitle{{AgentHarm}: A Benchmark for Measuring Harmfulness
  of {LLM} Agents}. In \bibinfo{booktitle}{\emph{The Thirteenth International
  Conference on Learning Representations (ICLR)}}.
\newblock
\urldef\tempurl%
\url{https://openreview.net/forum?id=AC5n7xHuR1}
\showURL{%
\tempurl}


\bibitem[Anil et~al\mbox{.}(2024)]%
        {anil2024manyshot}
\bibfield{author}{\bibinfo{person}{Cem Anil}, \bibinfo{person}{Esin Durmus},
  \bibinfo{person}{Nina Panickssery}, \bibinfo{person}{Mrinank Sharma},
  \bibinfo{person}{Joe Benton}, \bibinfo{person}{Sandipan Kundu},
  \bibinfo{person}{Joshua Batson}, {et~al\mbox{.}}}
  \bibinfo{year}{2024}\natexlab{}.
\newblock \showarticletitle{Many-shot Jailbreaking}. In
  \bibinfo{booktitle}{\emph{38th Conference on Neural Information Processing
  Systems (NeurIPS 2024)}}, Vol.~\bibinfo{volume}{37}.
  \bibinfo{pages}{129696--129742}.
\newblock
\href{https://doi.org/10.52202/079017-4121}{doi:\nolinkurl{10.52202/079017-4121}}


\bibitem[{Anthropic}(2024)]%
        {anthropic2024mcp}
\bibfield{author}{\bibinfo{person}{{Anthropic}}.}
  \bibinfo{year}{2024}\natexlab{}.
\newblock \bibinfo{title}{Model Context Protocol}.
\newblock \bibinfo{howpublished}{\url{https://modelcontextprotocol.io}}.
\newblock
\newblock
\shownote{Accessed on 23rd July 2026. Current version: 2024-11-05}.


\bibitem[Bhardwaj(2026)]%
        {bhardwaj2026skillfortify}
\bibfield{author}{\bibinfo{person}{Varun~Pratap Bhardwaj}.}
  \bibinfo{year}{2026}\natexlab{}.
\newblock \bibinfo{title}{{SkillFortify}: Formal Analysis and Supply Chain
  Security for Agentic {AI} Skills}.
\newblock
\href{https://doi.org/10.5281/zenodo.18787663}{doi:\nolinkurl{10.5281/zenodo.18787663}}


\bibitem[Bhatt et~al\mbox{.}(2025)]%
        {bhatt2025etdi}
\bibfield{author}{\bibinfo{person}{Manish Bhatt}, \bibinfo{person}{Vineeth~Sai
  Narajala}, {and} \bibinfo{person}{Idan Habler}.}
  \bibinfo{year}{2025}\natexlab{}.
\newblock \showarticletitle{ETDI: Mitigating Tool Squatting and Rug Pull
  Attacks in Model Context Protocol (MCP) by Using OAuth-Enhanced Tool
  Definitions and Policy-Based Access Control}. In
  \bibinfo{booktitle}{\emph{2025 Cyber Awareness and Research Symposium}}.
  \bibinfo{pages}{1--6}.
\newblock
\href{https://doi.org/10.1109/CARS67163.2025.11337310}{doi:\nolinkurl{10.1109/CARS67163.2025.11337310}}


\bibitem[{Cloud Security Alliance}(2026)]%
        {csa2026mcp}
\bibfield{author}{\bibinfo{person}{{Cloud Security Alliance}}.}
  \bibinfo{year}{2026}\natexlab{}.
\newblock \bibinfo{title}{{MCP} Security Crisis: Systemic Design Flaws in {AI}
  Agent Infrastructure}.
\newblock
  \bibinfo{howpublished}{\url{https://labs.cloudsecurityalliance.org/research/csa-research-note-mcp-security-crisis-20260504-csa-styled}}.
\newblock
\newblock
\shownote{Research Note. Accessed on 23rd July 2026}.


\bibitem[Cohen et~al\mbox{.}(2025)]%
        {cohen2025aiworm}
\bibfield{author}{\bibinfo{person}{Stav Cohen}, \bibinfo{person}{Ron Bitton},
  {and} \bibinfo{person}{Ben Nassi}.} \bibinfo{year}{2025}\natexlab{}.
\newblock \showarticletitle{Here Comes the AI Worm: Preventing the Propagation
  of Adversarial Self-Replicating Prompts Within GenAI Ecosystems}. In
  \bibinfo{booktitle}{\emph{Proceedings of the 2025 ACM SIGSAC Conference on
  Computer and Communications Security}} \emph{(\bibinfo{series}{CCS '25})}.
  \bibinfo{publisher}{Association for Computing Machinery},
  \bibinfo{address}{New York, NY, USA}, \bibinfo{pages}{3975--3989}.
\newblock
\href{https://doi.org/10.1145/3719027.3765196}{doi:\nolinkurl{10.1145/3719027.3765196}}


\bibitem[Debenedetti et~al\mbox{.}(2024)]%
        {debenedetti2024agentdojo}
\bibfield{author}{\bibinfo{person}{Edoardo Debenedetti}, \bibinfo{person}{Jie
  Zhang}, \bibinfo{person}{Mislav Balunovic}, \bibinfo{person}{Luca
  Beurer-Kellner}, \bibinfo{person}{Marc Fischer}, {and}
  \bibinfo{person}{Florian Tram{\`e}r}.} \bibinfo{year}{2024}\natexlab{}.
\newblock \showarticletitle{AgentDojo: A Dynamic Environment to Evaluate Prompt
  Injection Attacks and Defenses for {LLM} Agents}. In
  \bibinfo{booktitle}{\emph{The Thirty-eight Conference on Neural Information
  Processing Systems (NeurIPS) Datasets and Benchmarks Track}}.
\newblock
\urldef\tempurl%
\url{https://openreview.net/forum?id=m1YYAQjO3w}
\showURL{%
\tempurl}


\bibitem[Fu et~al\mbox{.}(2024)]%
        {fu2024imprompter}
\bibfield{author}{\bibinfo{person}{Xiaohan Fu}, \bibinfo{person}{Shuheng Li},
  \bibinfo{person}{Zihan Wang}, \bibinfo{person}{Yihao Liu},
  \bibinfo{person}{Rajesh~K. Gupta}, \bibinfo{person}{Taylor Berg-Kirkpatrick},
  {and} \bibinfo{person}{Earlence Fernandes}.} \bibinfo{year}{2024}\natexlab{}.
\newblock \bibinfo{title}{Imprompter: Tricking LLM Agents into Improper Tool
  Use}.
\newblock
\urldef\tempurl%
\url{https://arxiv.org/abs/2410.14923}
\showURL{%
\tempurl}


\bibitem[Greshake et~al\mbox{.}(2023)]%
        {greshake2023indirect}
\bibfield{author}{\bibinfo{person}{Kai Greshake}, \bibinfo{person}{Sahar
  Abdelnabi}, \bibinfo{person}{Shailesh Mishra}, \bibinfo{person}{Christoph
  Endres}, \bibinfo{person}{Thorsten Holz}, {and} \bibinfo{person}{Mario
  Fritz}.} \bibinfo{year}{2023}\natexlab{}.
\newblock \showarticletitle{Not What You've Signed Up For: Compromising
  Real-World {LLM}-Integrated Applications with Indirect Prompt Injection}. In
  \bibinfo{booktitle}{\emph{Proceedings of the 16th ACM Workshop on Artificial
  Intelligence and Security (AISec)}}. \bibinfo{publisher}{ACM},
  \bibinfo{address}{New York, NY, USA}.
\newblock
\href{https://doi.org/10.1145/3605764.3623985}{doi:\nolinkurl{10.1145/3605764.3623985}}


\bibitem[Guo et~al\mbox{.}(2026)]%
        {guo2025mcpxkit}
\bibfield{author}{\bibinfo{person}{Yongjian Guo}, \bibinfo{person}{Puzhuo Liu},
  \bibinfo{person}{Wanlun Ma}, \bibinfo{person}{Zehang Deng},
  \bibinfo{person}{Xiaogang Zhu}, \bibinfo{person}{Peng Di},
  \bibinfo{person}{Xi Xiao}, {and} \bibinfo{person}{Sheng Wen}.}
  \bibinfo{year}{2026}\natexlab{}.
\newblock \showarticletitle{{MCPXkit}: The Unified Toolkit for Analyzing Model
  Context Protocol Security}.
\newblock \bibinfo{journal}{\emph{IEEE Transactions on Dependable and Secure
  Computing}} \bibinfo{volume}{23}, \bibinfo{number}{5} (\bibinfo{year}{2026}),
  \bibinfo{pages}{9796--9811}.
\newblock
\href{https://doi.org/10.1109/TDSC.2026.3695553}{doi:\nolinkurl{10.1109/TDSC.2026.3695553}}


\bibitem[Hou et~al\mbox{.}(2026)]%
        {hou2025mcp}
\bibfield{author}{\bibinfo{person}{Xinyi Hou}, \bibinfo{person}{Yanjie Zhao},
  \bibinfo{person}{Shenao Wang}, {and} \bibinfo{person}{Haoyu Wang}.}
  \bibinfo{year}{2026}\natexlab{}.
\newblock \showarticletitle{Model Context Protocol (MCP): Landscape, Security
  Threats, and Future Research Directions}.
\newblock \bibinfo{journal}{\emph{ACM Trans. Softw. Eng. Methodol.}}
  (\bibinfo{date}{Feb.} \bibinfo{year}{2026}).
\newblock
\showISSN{1049-331X}
\href{https://doi.org/10.1145/3796519}{doi:\nolinkurl{10.1145/3796519}}


\bibitem[{Invariant Labs}(2025)]%
        {invariant2025tpa}
\bibfield{author}{\bibinfo{person}{{Invariant Labs}}.}
  \bibinfo{year}{2025}\natexlab{}.
\newblock \bibinfo{title}{{MCP} Security Notification: Tool Poisoning Attacks}.
\newblock
  \bibinfo{howpublished}{\url{https://invariantlabs.ai/blog/mcp-security-notification-tool-poisoning-attacks}}.
\newblock


\bibitem[Ji et~al\mbox{.}(2025)]%
        {ji2025ipidefensesok}
\bibfield{author}{\bibinfo{person}{Zimo Ji}, \bibinfo{person}{Xunguang Wang},
  \bibinfo{person}{Zongjie Li}, \bibinfo{person}{Pingchuan Ma},
  \bibinfo{person}{Yudong Gao}, \bibinfo{person}{Daoyuan Wu},
  \bibinfo{person}{Xincheng Yan}, \bibinfo{person}{Tian Tian}, {and}
  \bibinfo{person}{Shuai Wang}.} \bibinfo{year}{2025}\natexlab{}.
\newblock \bibinfo{title}{Taxonomy, Evaluation and Exploitation of IPI-Centric
  LLM Agent Defense Frameworks}.
\newblock
\urldef\tempurl%
\url{https://arxiv.org/abs/2511.15203}
\showURL{%
\tempurl}


\bibitem[Jiang et~al\mbox{.}(2026)]%
        {jiang2026capsafe}
\bibfield{author}{\bibinfo{person}{Xiaochong Jiang}, \bibinfo{person}{Shiqi
  Yang}, \bibinfo{person}{Ziwei Li}, \bibinfo{person}{Lifei Liu},
  \bibinfo{person}{Haoran Yu}, {and} \bibinfo{person}{Yichen Liu}.}
  \bibinfo{year}{2026}\natexlab{}.
\newblock \showarticletitle{{ChainCaps}: Composition-Safe Tool-Using Agents via
  Monotonic Capability Attenuation}. In \bibinfo{booktitle}{\emph{Proceedings
  of the Second Workshop on Agents in the Wild: Safety, Security, and Beyond
  (AIWILD @ ICML 2026)}}.
\newblock
\urldef\tempurl%
\url{https://arxiv.org/abs/2605.26542}
\showURL{%
\tempurl}


\bibitem[Kim et~al\mbox{.}(2026)]%
        {kim2026sok}
\bibfield{author}{\bibinfo{person}{Juhee Kim}, \bibinfo{person}{Wenbo Guo},
  {and} \bibinfo{person}{Dawn Song}.} \bibinfo{year}{2026}\natexlab{}.
\newblock \showarticletitle{{SoK}: Attack and Defense Landscape of Agentic {AI}
  Systems}. In \bibinfo{booktitle}{\emph{35th USENIX Security Symposium (USENIX
  Security 26)}}.
\newblock
\urldef\tempurl%
\url{https://www.usenix.org/conference/usenixsecurity26/presentation/kim-juhee-agentic}
\showURL{%
\tempurl}


\bibitem[Kim et~al\mbox{.}(2025)]%
        {kim2025manyshot}
\bibfield{author}{\bibinfo{person}{Sangyeop Kim}, \bibinfo{person}{Yohan Lee},
  \bibinfo{person}{Yongwoo Song}, {and} \bibinfo{person}{Kimin Lee}.}
  \bibinfo{year}{2025}\natexlab{}.
\newblock \showarticletitle{What Really Matters in Many-Shot Attacks? An
  Empirical Study of Long-Context Vulnerabilities in {LLMs}}. In
  \bibinfo{booktitle}{\emph{Proceedings of the 63rd Annual Meeting of the
  Association for Computational Linguistics (ACL 2025)}}.
\newblock
\urldef\tempurl%
\url{https://aclanthology.org/2025.acl-long.101/}
\showURL{%
\tempurl}


\bibitem[Li et~al\mbox{.}(2026)]%
        {li2026mcpitp}
\bibfield{author}{\bibinfo{person}{Ruiqi Li}, \bibinfo{person}{Zhiqiang Wang},
  \bibinfo{person}{Yunhao Yao}, {and} \bibinfo{person}{Xiang-Yang Li}.}
  \bibinfo{year}{2026}\natexlab{}.
\newblock \bibinfo{title}{{MCP-ITP}: An Automated Framework for Implicit Tool
  Poisoning in {MCP}}.
\newblock
\urldef\tempurl%
\url{https://arxiv.org/abs/2601.07395}
\showURL{%
\tempurl}


\bibitem[Li and Gao(2026)]%
        {li2025dsnmcp}
\bibfield{author}{\bibinfo{person}{Xiaofan Li} {and} \bibinfo{person}{Xing
  Gao}.} \bibinfo{year}{2026}\natexlab{}.
\newblock \showarticletitle{A First Look at the Security Issues in the Model
  Context Protocol Ecosystem}. In \bibinfo{booktitle}{\emph{2026 56th Annual
  IEEE/IFIP International Conference on Dependable Systems and Networks
  (DSN)}}. \bibinfo{publisher}{IEEE}, \bibinfo{pages}{379--392}.
\newblock
\href{https://doi.org/10.1109/DSN69566.2026.00046}{doi:\nolinkurl{10.1109/DSN69566.2026.00046}}


\bibitem[Liu et~al\mbox{.}(2026)]%
        {liu2026manuallies}
\bibfield{author}{\bibinfo{person}{Shi Liu}, \bibinfo{person}{Xuehai Tang},
  \bibinfo{person}{Xikang Yang}, \bibinfo{person}{Liang Lin},
  \bibinfo{person}{Biyu Zhou}, \bibinfo{person}{Wenjie Xiao}, {and}
  \bibinfo{person}{Wantao Liu}.} \bibinfo{year}{2026}\natexlab{}.
\newblock \showarticletitle{When the Manual Lies: A Realistic Benchmark to
  Evaluate {MCP} Poisoning Attacks for LLM Agents}. In
  \bibinfo{booktitle}{\emph{2026 29th International Conference on Computer
  Supported Cooperative Work in Design (CSCWD)}}. \bibinfo{publisher}{IEEE},
  \bibinfo{pages}{2090--2095}.
\newblock
\href{https://doi.org/10.1109/CSCWD68734.2026.11582512}{doi:\nolinkurl{10.1109/CSCWD68734.2026.11582512}}


\bibitem[{National Security Agency (NSA) / Cybersecurity and Infrastructure
  Security Agency}(2026)]%
        {nsa2025mcp}
\bibfield{author}{\bibinfo{person}{{National Security Agency (NSA) /
  Cybersecurity and Infrastructure Security Agency}}.}
  \bibinfo{year}{2026}\natexlab{}.
\newblock \bibinfo{title}{Model Context Protocol ({MCP}): Security Design
  Considerations for AI-Driven Automation}.
\newblock
  \bibinfo{howpublished}{\url{https://www.nsa.gov/Portals/75/documents/Cybersecurity/CSI_MCP_SECURITY.pdf}}.
\newblock
\newblock
\shownote{Accessed on 23rd July 2026}.


\bibitem[{OWASP Foundation}(2025)]%
        {owasp2025mcp03}
\bibfield{author}{\bibinfo{person}{{OWASP Foundation}}.}
  \bibinfo{year}{2025}\natexlab{}.
\newblock \bibinfo{title}{{OWASP MCP} Top 10 --- {MCP03}:2025 Tool Poisoning}.
\newblock
  \bibinfo{howpublished}{\url{https://owasp.org/www-project-mcp-top-10/}}.
\newblock


\bibitem[{OX Security}(2026)]%
        {ox2026mcp}
\bibfield{author}{\bibinfo{person}{{OX Security}}.}
  \bibinfo{year}{2026}\natexlab{}.
\newblock \bibinfo{title}{The Mother of All {AI} Supply Chains: Critical,
  Systemic Vulnerability at the Core of {Anthropic's MCP}}.
\newblock
  \bibinfo{howpublished}{\url{https://www.ox.security/blog/the-mother-of-all-ai-supply-chains-critical-systemic-vulnerability-at-the-core-of-the-mcp/}}.
\newblock
\newblock
\shownote{Published 15 April 2026. Accessed on 23rd July 2026}.


\bibitem[Pirch et~al\mbox{.}(2026)]%
        {pirch2026osagent}
\bibfield{author}{\bibinfo{person}{Lukas Pirch}, \bibinfo{person}{Micha
  Horlboge}, \bibinfo{person}{Patrick Großmann}, \bibinfo{person}{Syeda~Mahnur
  Asif}, \bibinfo{person}{Klim Kireev}, \bibinfo{person}{Thorsten Holz}, {and}
  \bibinfo{person}{Konrad Rieck}.} \bibinfo{year}{2026}\natexlab{}.
\newblock \bibinfo{title}{Toward Securing AI Agents Like Operating Systems}.
\newblock
\urldef\tempurl%
\url{https://arxiv.org/abs/2605.14932}
\showURL{%
\tempurl}


\bibitem[{Python Packaging Authority}(2026)]%
        {pypi2026stats}
\bibfield{author}{\bibinfo{person}{{Python Packaging Authority}}.}
  \bibinfo{year}{2026}\natexlab{}.
\newblock \bibinfo{title}{{PyPI} Download Statistics}.
\newblock \bibinfo{howpublished}{\url{https://pypistats.org}}.
\newblock
\newblock
\shownote{Accessed on 23rd July 2026}.


\bibitem[Rostamzadeh et~al\mbox{.}(2026)]%
        {rostamzadeh2026mcpdpt}
\bibfield{author}{\bibinfo{person}{Mehrdad Rostamzadeh},
  \bibinfo{person}{Sidhant Narula}, \bibinfo{person}{Nahom Birhan},
  \bibinfo{person}{Mohammad Ghasemigol}, {and} \bibinfo{person}{Daniel
  Takabi}.} \bibinfo{year}{2026}\natexlab{}.
\newblock \bibinfo{title}{MCP-DPT: A Defense-Placement Taxonomy and Coverage
  Analysis for Model Context Protocol Security}.
\newblock
\urldef\tempurl%
\url{https://arxiv.org/abs/2604.07551}
\showURL{%
\tempurl}


\bibitem[Russinovich et~al\mbox{.}(2025)]%
        {russinovich2024crescendo}
\bibfield{author}{\bibinfo{person}{Mark Russinovich}, \bibinfo{person}{Ahmed
  Salem}, {and} \bibinfo{person}{Ronen Eldan}.}
  \bibinfo{year}{2025}\natexlab{}.
\newblock \showarticletitle{Great, Now Write an Article About That: The
  Crescendo Multi-Turn {LLM} Jailbreak Attack}. In
  \bibinfo{booktitle}{\emph{34th USENIX Security Symposium (USENIX Security
  25)}}.
\newblock
\urldef\tempurl%
\url{https://www.usenix.org/conference/usenixsecurity25/presentation/russinovich}
\showURL{%
\tempurl}


\bibitem[Schulhoff et~al\mbox{.}(2023)]%
        {schulhoff2023hackaprompt}
\bibfield{author}{\bibinfo{person}{Sander Schulhoff}, \bibinfo{person}{Jeremy
  Pinto}, \bibinfo{person}{Anaum Khan}, \bibinfo{person}{Louis-Fran\c{c}ois
  Bouchard}, \bibinfo{person}{Chenglei Si}, \bibinfo{person}{Svetlina Anati},
  \bibinfo{person}{Valen Tagliabue}, \bibinfo{person}{Anson Kost},
  \bibinfo{person}{Christopher Carnahan}, \bibinfo{person}{Jordan Boyd-Graber},
  {et~al\mbox{.}}} \bibinfo{year}{2023}\natexlab{}.
\newblock \showarticletitle{Ignore This Title and {HackAPrompt}: Exposing
  Systemic Vulnerabilities of {LLMs} Through a Global Prompt Hacking
  Competition}. In \bibinfo{booktitle}{\emph{Proceedings of the 2023 Conference
  on Empirical Methods in Natural Language Processing (EMNLP)}}.
  \bibinfo{publisher}{Association for Computational Linguistics},
  \bibinfo{address}{Singapore}, \bibinfo{pages}{4945--4977}.
\newblock
\href{https://doi.org/10.18653/v1/2023.emnlp-main.302}{doi:\nolinkurl{10.18653/v1/2023.emnlp-main.302}}


\bibitem[Wang et~al\mbox{.}(2026c)]%
        {wang2026adaptools}
\bibfield{author}{\bibinfo{person}{Che Wang}, \bibinfo{person}{Jiaming Zhang},
  \bibinfo{person}{Ziqi Zhang}, \bibinfo{person}{Zijie Wang},
  \bibinfo{person}{Yinghui Wang}, \bibinfo{person}{Jianbo Gao},
  \bibinfo{person}{Tao Wei}, \bibinfo{person}{Zhong Chen}, {and}
  \bibinfo{person}{Wei Yang~Bryan Lim}.} \bibinfo{year}{2026}\natexlab{c}.
\newblock \bibinfo{title}{{AdapTools}: Adaptive Tool-based Indirect Prompt
  Injection Attacks on Agentic {LLMs}}.
\newblock
\urldef\tempurl%
\url{https://arxiv.org/abs/2602.20720}
\showURL{%
\tempurl}


\bibitem[Wang et~al\mbox{.}(2026a)]%
        {wang2025mindguard}
\bibfield{author}{\bibinfo{person}{Zhiqiang Wang}, \bibinfo{person}{Haohua Du},
  \bibinfo{person}{Guanquan Shi}, \bibinfo{person}{Junyang Zhang},
  \bibinfo{person}{HaoRan Cheng}, \bibinfo{person}{Yunhao Yao},
  \bibinfo{person}{Kaiwen Guo}, {and} \bibinfo{person}{Xiang-Yang Li}.}
  \bibinfo{year}{2026}\natexlab{a}.
\newblock \bibinfo{title}{{MindGuard}: Intrinsic Decision Inspection for
  Securing LLM Agents Against Metadata Poisoning}.
\newblock
\urldef\tempurl%
\url{https://arxiv.org/abs/2508.20412}
\showURL{%
\tempurl}


\bibitem[Wang et~al\mbox{.}(2026b)]%
        {wang2025mcptox}
\bibfield{author}{\bibinfo{person}{Zhiqiang Wang}, \bibinfo{person}{Yichao
  Gao}, \bibinfo{person}{Yanting Wang}, \bibinfo{person}{Suyuan Liu},
  \bibinfo{person}{Haifeng Sun}, \bibinfo{person}{Haoran Cheng},
  \bibinfo{person}{Guanquan Shi}, \bibinfo{person}{Haohua Du}, {and}
  \bibinfo{person}{Xiang-Yang Li}.} \bibinfo{year}{2026}\natexlab{b}.
\newblock \showarticletitle{{MCPTox}: A Benchmark for Tool Poisoning on
  Real-World {MCP} Servers}. In \bibinfo{booktitle}{\emph{Proceedings of the
  AAAI Conference on Artificial Intelligence}}, Vol.~\bibinfo{volume}{40}.
  \bibinfo{publisher}{AAAI Press}, \bibinfo{pages}{35811--35819}.
\newblock
\href{https://doi.org/10.1609/aaai.v40i42.40895}{doi:\nolinkurl{10.1609/aaai.v40i42.40895}}


\bibitem[Xie et~al\mbox{.}(2026)]%
        {xie2026stored}
\bibfield{author}{\bibinfo{person}{Yuanbo Xie}, \bibinfo{person}{Wenlei Zhu},
  \bibinfo{person}{Tianyun Liu}, \bibinfo{person}{Yingjie Zhang},
  \bibinfo{person}{Suchen Liu}, \bibinfo{person}{Yulin Li},
  \bibinfo{person}{Liya Su}, {and} \bibinfo{person}{Tingwen Liu}.}
  \bibinfo{year}{2026}\natexlab{}.
\newblock \bibinfo{title}{What If Prompt Injection Never Left? Rethinking Agent
  Security through Cross-Session Stored Prompt Injection}.
\newblock
\urldef\tempurl%
\url{https://arxiv.org/abs/2606.04425}
\showURL{%
\tempurl}


\bibitem[Yang et~al\mbox{.}(2026)]%
        {yang2026sok}
\bibfield{author}{\bibinfo{person}{Shiqi Yang}, \bibinfo{person}{Wenting Yang},
  \bibinfo{person}{Xiaochong Jiang}, \bibinfo{person}{Cheng Ji}, {and}
  \bibinfo{person}{Yichen Liu}.} \bibinfo{year}{2026}\natexlab{}.
\newblock \showarticletitle{{SoK}: A Taxonomy of Attack Vectors and Defense
  Strategies for Agentic Supply Chain Runtime}. In
  \bibinfo{booktitle}{\emph{ICLR 2026 Workshop on AI for Mechanism Design and
  Strategic Decision Making (AIMS)}}.
\newblock
\urldef\tempurl%
\url{https://openreview.net/forum?id=RpWthPyigJ}
\showURL{%
\tempurl}


\bibitem[Ye et~al\mbox{.}(2024)]%
        {ye2024toolsword}
\bibfield{author}{\bibinfo{person}{Junjie Ye}, \bibinfo{person}{Sixian Li},
  \bibinfo{person}{Guanyu Li}, \bibinfo{person}{Caishuang Huang},
  \bibinfo{person}{Songyang Gao}, \bibinfo{person}{Yilong Wu},
  \bibinfo{person}{Qi Zhang}, \bibinfo{person}{Tao Gui}, {and}
  \bibinfo{person}{Xuanjing Huang}.} \bibinfo{year}{2024}\natexlab{}.
\newblock \showarticletitle{{ToolSword}: Unveiling Safety Issues of Large
  Language Models in Tool Learning Across Three Stages}. In
  \bibinfo{booktitle}{\emph{Proceedings of the 62nd Annual Meeting of the
  Association for Computational Linguistics (Volume 1: Long Papers)}}.
  \bibinfo{publisher}{Association for Computational Linguistics},
  \bibinfo{address}{Bangkok, Thailand}, \bibinfo{pages}{2181--2211}.
\newblock
\href{https://doi.org/10.18653/v1/2024.acl-long.119}{doi:\nolinkurl{10.18653/v1/2024.acl-long.119}}


\bibitem[Yuan et~al\mbox{.}(2024)]%
        {yuan2024cipherchat}
\bibfield{author}{\bibinfo{person}{Youliang Yuan}, \bibinfo{person}{Wenxiang
  Jiao}, \bibinfo{person}{Wenxuan Wang}, \bibinfo{person}{Jen-tse Huang},
  \bibinfo{person}{Pinjia He}, \bibinfo{person}{Shuming Shi}, {and}
  \bibinfo{person}{Zhaopeng Tu}.} \bibinfo{year}{2024}\natexlab{}.
\newblock \showarticletitle{{GPT-4} Is Too Smart To Be Safe: Stealthy Chat with
  {LLMs} via Cipher}. In \bibinfo{booktitle}{\emph{International Conference on
  Learning Representations (ICLR)}}. \bibinfo{address}{Vienna, Austria}.
\newblock
\urldef\tempurl%
\url{https://openreview.net/forum?id=MbfAK4s61A}
\showURL{%
\tempurl}


\bibitem[Zhan et~al\mbox{.}(2024)]%
        {zhan2024injecagent}
\bibfield{author}{\bibinfo{person}{Qiusi Zhan}, \bibinfo{person}{Zhixiang
  Liang}, \bibinfo{person}{Zifan Ying}, {and} \bibinfo{person}{Daniel Kang}.}
  \bibinfo{year}{2024}\natexlab{}.
\newblock \showarticletitle{{InjecAgent}: Benchmarking Indirect Prompt
  Injections in Tool-Integrated Large Language Model Agents}. In
  \bibinfo{booktitle}{\emph{Findings of the Association for Computational
  Linguistics: {ACL} 2024}}. \bibinfo{publisher}{Association for Computational
  Linguistics}, \bibinfo{address}{Bangkok, Thailand},
  \bibinfo{pages}{10471--10506}.
\newblock
\href{https://doi.org/10.18653/v1/2024.findings-acl.624}{doi:\nolinkurl{10.18653/v1/2024.findings-acl.624}}


\bibitem[Zhang et~al\mbox{.}(2026)]%
        {msb2026}
\bibfield{author}{\bibinfo{person}{Dongsen Zhang}, \bibinfo{person}{Zekun Li},
  \bibinfo{person}{Xu Luo}, \bibinfo{person}{Xuannan Liu},
  \bibinfo{person}{Pei~Pei Li}, {and} \bibinfo{person}{Wenjun Xu}.}
  \bibinfo{year}{2026}\natexlab{}.
\newblock \showarticletitle{{MCP} Security Bench ({MSB}): Benchmarking Attacks
  Against Model Context Protocol in {LLM} Agents}. In
  \bibinfo{booktitle}{\emph{The Fourteenth International Conference on Learning
  Representations (ICLR)}}.
\newblock
\urldef\tempurl%
\url{https://openreview.net/forum?id=irxxkFMrry}
\showURL{%
\tempurl}


\bibitem[Zhang et~al\mbox{.}(2025)]%
        {zhang2025asb}
\bibfield{author}{\bibinfo{person}{Hanrong Zhang}, \bibinfo{person}{Jingyuan
  Huang}, \bibinfo{person}{Kai Mei}, \bibinfo{person}{Yifei Yao},
  \bibinfo{person}{Zhenting Wang}, \bibinfo{person}{Chenlu Zhan},
  \bibinfo{person}{Hongwei Wang}, {and} \bibinfo{person}{Yongfeng Zhang}.}
  \bibinfo{year}{2025}\natexlab{}.
\newblock \showarticletitle{Agent Security Bench ({ASB}): Formalizing and
  Benchmarking Attacks and Defenses in {LLM}-based Agents}. In
  \bibinfo{booktitle}{\emph{International Conference on Learning
  Representations (ICLR)}}. \bibinfo{publisher}{OpenReview.net},
  \bibinfo{address}{Singapore}.
\newblock
\urldef\tempurl%
\url{https://openreview.net/forum?id=V4y0CpX4hK}
\showURL{%
\tempurl}


\bibitem[Zhou et~al\mbox{.}(2026)]%
        {zhou2026mcpauth}
\bibfield{author}{\bibinfo{person}{Huijun Zhou}, \bibinfo{person}{Xiaohan
  Zhang}, \bibinfo{person}{Haozhe Zhang}, \bibinfo{person}{Haoyang Zhang},
  \bibinfo{person}{Mi Zhang}, {and} \bibinfo{person}{Min Yang}.}
  \bibinfo{year}{2026}\natexlab{}.
\newblock \bibinfo{title}{A First Measurement Study on Authentication Security
  in Real-World Remote {MCP} Servers}.
\newblock
\urldef\tempurl%
\url{https://arxiv.org/abs/2605.22333}
\showURL{%
\tempurl}


\end{thebibliography}

\section*{Declaration on the Use of Generative AI}
\label{sec:genai-disclosure}
In accordance with the ACM Policy on Authorship and the AISec submission requirements, the use of generative AI in the preparation of this work is disclosed here.
Claude Code (Anthropic) was used for implementation support in the \mcpvenom harness and the accompanying analysis scripts, and for copy-editing of the manuscript; all such output was reviewed, executed, and validated by the authors, who retain full responsibility for the content.
No claim, table, figure, or citation was accepted without verification against the released artifacts.
Language models are otherwise the object of measurement rather than an instrument of it: the dual oracle is deterministic, resting on canary-server request logs and pattern matching over request bodies, and thus no attack-success label depends on model judgement.

\end{document}